\documentclass[prx,twocolumn,groupeaddress]{revtex4-1}

\usepackage{amsmath}
\usepackage{amssymb}
\usepackage{xspace}
\usepackage{graphicx}
\usepackage{grffile}
\usepackage{nicefrac}
\usepackage{array}
\usepackage{color}

\graphicspath{{figs/}}

\newcommand{\ket}[1]{\left.\left|{#1} \right\rangle\right.}
\newcolumntype{L}[1]{>{\raggedright\let\newline\\\arraybackslash\hspace{0pt}}m{#1}}
\newcolumntype{C}[1]{>{\centering\let\newline\\\arraybackslash\hspace{0pt}}m{#1}}
\newcolumntype{R}[1]{>{\raggedleft\let\newline\\\arraybackslash\hspace{0pt}}m{#1}}
\newcommand{\bw}[1]{\raisebox{1.5ex}[-1.5ex]{#1}}
\def\bk{{\bf k}}
\def\br{{\bf r}}
\def\bR{{\bf R}}

\begin{document}

\title{From Basic Properties to the Mott Design of Correlated Delafossites}
\author{Frank Lechermann}
\affiliation{
\small{European XFEL, Holzkoppel 4, 22869 Schenefeld, Germany}, and
\small{Center for Computational Quantum Physics, Flatiron Institute, 
162 5th Avenue, New York, New York 10010, USA}}

\begin{abstract}
The natural-heterostructure concept realized in delafossites highlights these 
layered oxides. While metallic, band- or Mott-insulating character may be associated 
with individual layers, inter-layer coupling still plays a decisive role. We  
review the correlated electronic structure of PdCoO$_2$, PdCrO$_2$ and AgCrO$_2$,  
showing that layer-entangled electronic states can deviate from standard 
classifications of interacting systems. This finding opens up possibilities 
for materials design in a subtle Mott-critical regime.
Manipulated Hidden-Mott physics, correlation-induced semimetallicity, or Dirac/flat-band 
dispersions in a Mott background are emerging features. Together with achievements in 
the experimental preparation, this inaugurates an exciting research field in the arena 
of correlated materials.
\end{abstract}

\maketitle

\section*{Introduction}
A common statement on the research on strongly correlated materials, especially from a
theory perspective, refers to the fact that the body of compounds 
that fall into this category is rather small, and hence the relevance and impact in 
general materials science scales accordingly. While there has been some truth in this
view, there are also obvious facts that argue against. 
First undoubtedly, the properties of various correlated systems are singular. 
High-temperature superconductivity in cuprates or colossal magnetoresistance in 
manganites are only two extraordinary features that stand out in condensed matter 
physics. Moreover, even 'straightforward' materials properties, such as e.g. magnetism
in the solid state, are often relying on some effect of electronic correlation.
Second, times are changing and from an applicative viewpoint, understanding
and engineering electron correlation may be indispensable to make progress in battery
materials, thermoelectric devices, photovoltaic systems, data-storage media and 
other sorts of 'smart' materials. Against this background, we want to discuss in this 
overview the fascinating physics of a certain class of transition-metal (TM) oxides prone 
to correlation effects, namely the delafossites~\cite{fri73,rog13,sha71-1,sha71-2,sha71-3}. 
Importantly, this materials class is not only of interest due to its intrinsic basic 
properties, but also because of the potential for even more intriguing characteristics upon 
further design. 
\begin{figure}[t]
\begin{center}
\includegraphics*[width=8.25cm]{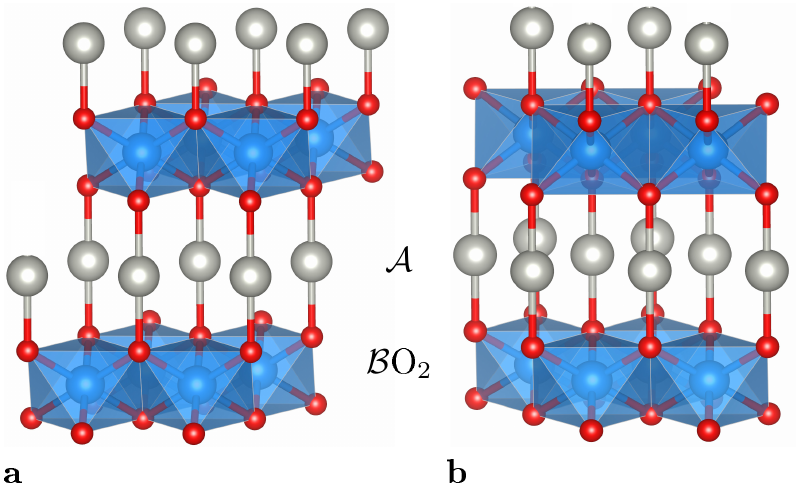}
\end{center}
\caption{
{\bf Delafossite ${\cal A}{\cal B}$O$_2$ crystal structure}. ({\bf a}) $R\bar{3}m$ and
({\bf b}) $P6_3/mmc$ symmetry. ${\cal A}$ (grey), ${\cal B}$ (blue) and O (red).
\label{fig:struc}}
\end{figure}

In order to theoretically investigate challenging systems with subtle electronic 
characteristics, an advanced framework is needed, capable of addressing electron
states from weak to strong correlation on an equal footing. Model-Hamiltonian approaches
may only be used at a later stage, when focussing on certain details of the complex
quantum problem. Density functional theory (DFT) in Kohn-Sham representation is proper
to describe the band formation from first principles, but will not be sufficient to 
account for relevant correlation effects. The combination of DFT with dynamical 
mean-field theory (DMFT), the so-called DFT+DMFT method~\cite{ani97,lic98,kot06}, 
is well suited for the problem, as it can account for site- and orbital-resolved Mott 
criticality at strong coupling as well as for mildly renormalized dispersions at weak 
coupling in a realistic setting.\\

Delafossites, named after the french crystallographer Gabriel Delafosse (1796-1878), 
are in fact known for quite some time. The first delafossite, the CuFeO$_2$ mineral, 
has been discovered~\cite{fri73} near Yekaterinburg, Russia, in 1873 and then 
re-discovered~\cite{rog13} near Bisbee(Arizona), USA, in 1913.
Since then, numerous compounds of the delafossite-oxide type ${\cal A}{\cal B}$O$_2$, 
where ${\cal A}$ and ${\cal B}$ denote different metallic elements, have been 
crystallized. 
The unique crystal structure (see Fig.~\ref{fig:struc}) consists of an alternate 
stacking of triangular ${\cal A}$ lattices and planes of edge-sharing ${\cal B}$O$_6$ 
octahedra along the 
$c$-axis, whereby these two different layer types are connected via oxygen in a so-called 
dumbbell position. There are two possible stacking scenarios, 
namely the more common rhombohedral one with $R\bar{3}m$ space-group symmetry and the 
hexagonal one giving rise to $P6_3/mmc$ symmetry. The metallic ions are in the formal
oxidation state ${\cal A}^+$ and ${\cal B}^{3+}$, respectively.

Delafossites are divided into a larger insulating and a smaller metallic class of compounds. 
In modern times, the $p$-type electrical conduction in the transparent CuAlO$_2$ 
insulator~\cite{kaw97} gained strongest interest. 
In a series of papers~\cite{sha71-1,sha71-2,sha71-3}, Shannon {\sl et al.}
in 1971 described the novel synthesis and single-crystal growth of several delafossites
with ${\cal A}$=Pd, Pt and Ag. Among those, there are oxides with exceptionally high electrical
conductivity at room temperature, e.g. PdCoO$_2$ and PtCoO$_2$, in combination
with an outstanding single-crystal purity. This combined feature found in a selected 
subgroup of delafossites has started to become an intense field of research
(see e.g. Refs.~\cite{dao17,mac17} for reviews). This elitist group of delafossites 
in terms of metallic properties, includes the PdCrO$_2$ compound, which, among 
further challenging physics, hosts Mott-insulating CrO$_2$ layers~\cite{noh14,lec18,sun20}.

This brings us to a very relevant aspect. The special delafossite architecture
gives rise to a natural heterostructure, in which individual layers may attain a distinct
character of their own. In most layered materials, e.g. cuprates, cobaltates, etc., 
there is usually one 'active' layer type and the remaining part mainly provides the glue. 
However in delafossites, e.g. the ${\cal A}$ layer can 
manage the metallic transport, while the ${\cal B}$O$_2$ layers account for the 
magnetic ordering. This not only entails exciting physical processes in the pure 
compound, but furthermore allows for a kind of 'meta oxide-heterostructure' physics 
upon additional (nano-)engineering. 

Since the family of delafossites encircles a vast number of compounds with many different 
physical aspects, we cannot reasonably-well cover all those in this comparatively brief 
treatise. Instead, the goal is to focus on the specific role of electron correlations in 
selected compounds with ${\cal A}$=Pd, Ag and ${\cal B}$=Co, Cr as well as an assessment of 
possible designing routes. For further discussion of delafossites we refer the reader to 
the exisiting body of literature, e.g. Refs.~\cite{dao17,mac17,hud09,ses98,she08} and 
references therein.\\

One usually refers to the term ``Mottness'' in order to mark the phenomenology of an electronic 
system either close to a Coulomb-interaction driven metal-to-insulator transition or already in
a Mott-insulating state. It most often includes e.g. abrupt changes in transport behavior, 
unique temperature dependencies and/or strong magnetic response due to the formation/existence of 
local moments. As a rule of thumb, the poorly screened Coulomb interaction exceeds 
(or is comparabale to) the relevant bandwidth in corresponding materials. In TM oxides, there are 
various cases that give rise to Mottness, which depend on certain properties of the TM ion and 
the actual crystal structure (see e.g. Ref.~\cite{ima98} and references therein for a detailed
review). 

When starting from a non-interacting metallic band structure, in conventional cases a given 
band manifold crossing the Fermi level is subject to Coulomb interactions that arise locally
from the TM site, i.e. associated with a Hubbard $U$. The size of the orbital space that defines 
this specific band manifold as well as its filling are crucial properties. 
The one-band/one-orbital limit, though most appreciated from a model perspective, is however 
very rarely realized in concrete materials. 
Further key property is the placement of the TM element in the periodic table, which regulates the 
TM$(d)$ vs. O$(2p)$ level position~\cite{zaa85}: 
strongly correlated early TM oxides are mostly of Mott-Hubbard 
type, whereas such late TM oxides are mostly of charge-transfer type. Moreover, the correlation 
signatures depend on the TM-element row. For instance, spin-orbit effect are often non-negligible 
in $4d$ and especially $5d$ oxides, whereby the corresponding $U$ values are usually smaller than 
in $3d$ oxides.

Then there are more detailed differentiations that arise from specific compound characteristics.
For instance, in orbital-selective Mott systems~\cite{ani02} only a subclass of the relevant 
orbitals become 
Mott critical. Explicit non-local correlation effects, e.g. on selected short-range bonds, 
play a key role in a few materials~\cite{bie05}. Doping with impurities can lead to novel behavior,
e.g. the appearance of site-selective Mott behavior where interaction-driven localization
tendencies occurs only on selected lattice sites~\cite{lec20-2}. Last but not least, 
so-called Hund metals (see e.g.~\cite{geo13} for a review) are in principle distant from a 
Mott-critical regime, but enable features of strong correlation based on an interplay of $U$ 
and the Hund's exchange $J_{\rm H}$. While Mottness is usually strongest for half-filled orbital 
manifolds, the latter ``Hundness'' is usually strongest for orbital manifolds with one electron(hole)
added to half filling.

Notably, electronic correlation in selected delafossites adds a further scenario to the list. 
Their specific layered structure can account for layer-selective Mott criticality that
may furthermore be connected to singular layer-to-layer coupling. This means that only certain
layers within the delafossite structure are in or close to a Mott state, whereas remaining layers
are well conducting. This kind of selectivity can on the other hand give rise to non-conventional
quantum states for electrons that want to move coherently throughout the complete system.

Before delving into the details of the correlated electronic structure of delafossites,
a reminder of the state-of-the-art DFT+DMFT approach is given in the following
section. Since this review is intended to focus on the materials and the
correlation-design aspect, that theory part will be rather brief. There are already various 
more extensive descriptions of DFT+DMFT, e.g.~\cite{geo04,kot06,lec18-2}.

\section*{Density functional theory plus\\ dynamical mean-field theory}
\subsection*{General formalism}
Describing the general many-body problem from weak to strong coupling in a condensed matter 
system within a first-principles(-like) manner is tough. A unique and well-defined solution
has not been given yet. Especially when one also wants to address materials 
science questions with larger unit cells and larger orbital manifolds, a solution presumably 
has to wait for much longer times. Approximate hybrid methods that divide the complex problem
into (coupled) subproblems of different significance have proven adequate to obtain good 
results beyond effective single-particle schemes. The DFT+DMFT technique is such a hybrid 
method.
\begin{figure}[b]
\begin{center}
\includegraphics*[width=8.5cm]{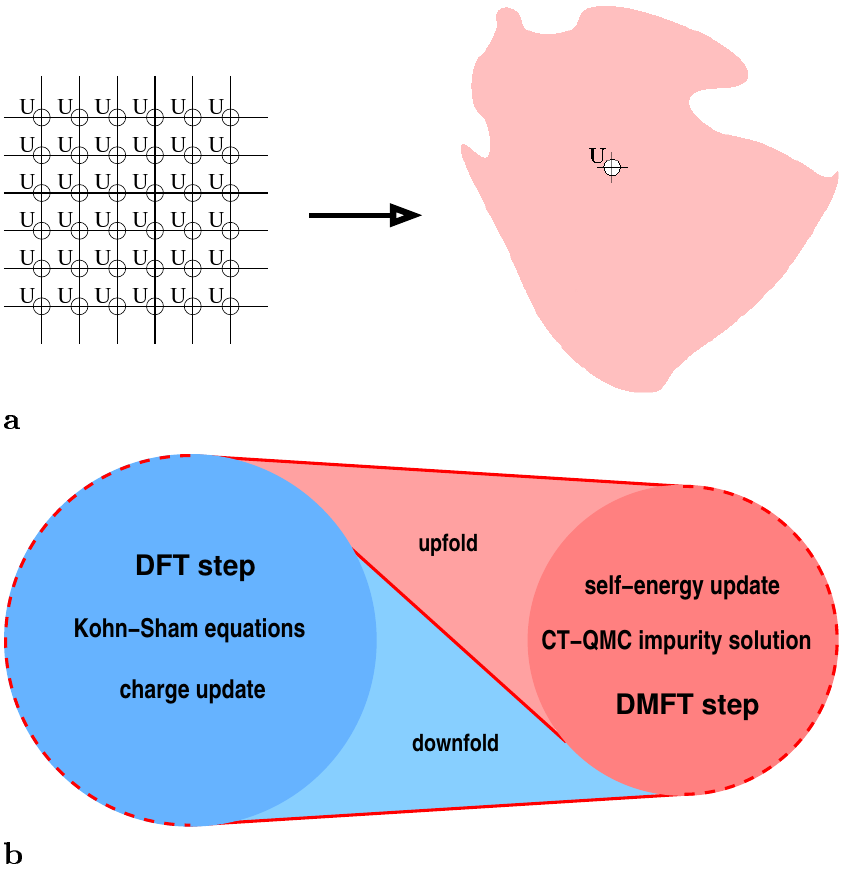}
\end{center}
\caption{{\bf Basics of the realistic-DMFT framework}. 
({\bf a}) Sketch of the DMFT mapping. An interacting electron system on a lattice
with a Coulomb repulsion $U$ on each site is mapped onto a single interacting site
in an energy-dependent bath. The latter is determined self-consistently in an iterative
cycle. ({\bf b}) State-of-the-art charge self-consistent DFT+DMFT loop 
(after~\cite{lec06}). The calculation usually starts from a self-consistent Kohn-Sham
solution. The correlated subspace is defined and the initial Weiss field ${\mathcal G}_0$
constructed. Afterwards, a single DMFT step is performed. The obtained
self-energies are upfolded and an updated charge density $n({\bf r})$ is computed. 
A new charge density implies a new Kohn-Sham potential, and a single new
Kohn-Sham step is performed, therefrom a new Weiss field is generated, etc..}
\label{fig:dmft}       
\end{figure}
We assume that the reader is familiar with DFT and we note that the term 'DFT' is understood 
throughtout the text as 'effective single-particle Kohn-Sham DFT'. While this discrimination 
is important, we here focus on practical calculations and shorten the abbrevation for 
readability matters.

The dynamical mean-field theory~\cite{met89,geo92} was invented for model Hamiltonians and is
appreciated as the many-body scheme with the best compromise between generality,
accuracy and performance. Just as Kohn-Sham DFT, also DMFT describes a mapping: from the 
problem of interacting lattice electrons onto the problem of a quantum impurity within a 
self-consistent energy-dependent bath, sketched in Fig.~\ref{fig:dmft}a. 
Key focus is on the one-particle Green's function, which for 
chemical potential $\mu$ and Hamiltonian $H(\bk)$ at wave vector $\bk$ reads
\begin{equation}
G(\bk,i\omega_{\rm n})=[i\omega_{\rm n}+\mu-H(\bk)-\Sigma(\bk,i\omega_{\rm n})]^{-1}\quad.
\end{equation}
Note that here, fermionic Matsubara frequencies 
$\omega_{\rm n}:=(2n+1)\pi T$ are employed to emphasize the treatment at finite 
temperature. The analytical continuation to real frequencies $\omega$ in actual calculations 
may e.g. be performed via the maximum entropy method (see e.g.~\cite{ima98,geo96,jar96} 
for more details). The self-energy $\Sigma$ describes the many-body part of
the problem, and hence, finding a good approximation is key to a physically sound
picture. In DFT, the self-energy is approximated, in essence, by the energy-independent 
sum of Hartree potential $v_{\rm H}$ and exchange-correlation potential 
$v_{xc}$ in the form of a simple forward-scattering term.
In DMFT, the local Green's function is approximated with the help of a $\bk$-independent 
but energy-dependent impurity self-energy $\Sigma_{\rm imp}(i\omega_{\rm n})$, i.e.
\begin{equation}
G_{\rm loc}^{\rm DMFT}(i\omega_{\rm n})=\sum_{\bk}[i\omega_{\rm n}+\mu-H(\bk)
-\Sigma_{\rm imp}(i\omega_{\rm n})]^{-1}\quad,\label{eq:lgreen}
\end{equation}
whereby the corresponding impurity problem is defined via
\begin{equation}
\Sigma_{\rm imp}(i\omega_{\rm n})={\mathcal G}_0(i\omega_{\rm n})^{-1}
-G_{\rm imp}(i\omega_{\rm n})^{-1}\quad.
\label{eq:imp}
\end{equation}
The Weiss field ${\mathcal G}_0(i\omega_{\rm n})$ is a function of the local Hamiltonian 
(expressed within a localized basis). The important DMFT self-consistency condition 
implies $G_{\rm imp}=G_{\rm loc}^{\rm DMFT}$ and is usually achieved within a loop,
just as the Kohn-Sham cycle. From an initial version of ${\mathcal G}_0$, 
the self-energy $\Sigma_{\rm imp}$ is determined and with the use 
of (\ref{eq:lgreen}),(\ref{eq:imp})
a new ${\mathcal G}_0$ extracted, and so on. The hard part consists in solving the 
quantum-impurity problem to obtain $G_{\rm imp}$ for a given Weiss field. 
So-called 'solvers' based e.g., on quantum Monte Carlo, 
Exact Diagonalization, etc. are employed for that task. Note that local-interaction diagrams 
are included to all orders in this non-perturbative theory. The vital energy dependence of 
the Weiss field ensures the qualitatively correct description of low-energy quasiparticle 
(QP) features as well as high-energy incoherent (Hubbard) excitations. Extensions to overcome 
the restriction to a local self-energy, e.g. via cluster schemes, are available,
but will not be pursued here.

We now come back to the portioning into subspaces within DFT+DMFT. Concerning DFT for a few
hundred sites, there is no issue for conventional exchange-correlation functionals and 
one may apply that approach to the complete electronic Hilbert space. This ensures a reliable
description of the bonding, band formation and screening properties in a given material.
On the other hand, DMFT as a manifest many-body scheme is not applicable to hundreds of sites.
Furthermore and very importantly, there are further issues to the use of pure DMFT in a
concrete materials context. First, DMFT builds up on the physics of interactions in orbitals
with reasonably local character, i.e. a Hubbard-model-like scenario. However, such a scenario
is not straightforwardly suitable for e.g. dominant $s$ and $p$ electron states. Second, DMFT
is designed to provide proper access to the self-energy $\Sigma$, but not to derive general
hoppings $t$. In other words, a full-monty DMFT starting from atomic Coulomb potentials on
a given lattice is just not the conventional modus operandi. Note that there are ideas to 
use DMFT for quantum chemistry problems in a direct manner~\cite{zgi11,lin11}, but it is still
a long way to condensed matter materials science. Therefore, putting materials-oriented DMFT 
into practise is currently best done by allocating a restricted Hilbert subspace, i.e.
the so-called correlated subspace.

The correlated subspace is understood as a quantum-numbered real-space region where 
correlated electrons reside. Note that this subspace is not uniquely defined, but is a matter
of choice in a concrete materials problem. For instance, in the case of an early 
transition-metal oxide, like e.g. SrVO$_3$ or V$_2$O$_3$, it may consist of the low-energy 
$t_{2g}$ orbitals of the TM site(s).
Within the correlated subspace, a multi-orbital interacting Hamiltonian is applied. 
More concretely, the corresponding Hamiltonian terms are explicitly 
exploited in the impurity solver. The Hamiltonian is usually of generalized Hubbard 
type, with local interaction parameters based on the Coulumb integral $U$ and the Hund 
exchange $J_{\rm H}$. Those are either chosen by hand or computed ab-initio, e.g via the 
constrained random-phase approximation~\cite{ary04}.

Key interfaces of the complete DFT+DMFT self-consistency 
cycle~\cite{sav01,min05,pou07,gri12} (cf. Fig.~\ref{fig:dmft}b) are marked by the 
downfolding of the full-problem Bloch (bl) space to the correlated subspace, 
and the upfolding of the DMFT self-energy back to the full space. In terms of the 
Green's function $G$ and the self-energy $\Sigma$, for sites $\bR$, local orbitals $mm'$ 
and band indices $\nu\nu'$ this reads formula-wise
\begin{eqnarray}
&&G^{\bR,{\rm imp}}_{mm'}(i\omega_{\rm n})=
\hspace*{-0.2cm}\sum_{\bk,(\nu\nu')\in {\cal W}}\hspace*{-0.5cm}
\bar{P}^{\bR}_{m\nu}(\bk)\,G^{\rm bl}_{\nu\nu'}(\bk,i\omega_{\rm n})\,
\bar{P}^{\bR*}_{\nu' m'}(\bk)\;,\label{eq:g_limband}\\
&&\Delta\Sigma^{\rm bl}_{\nu\nu'}(\bk,i\omega_{\rm n})=
\hspace*{-0.2cm}\sum_{\bR,mm'}\hspace*{-0.3cm}
\bar{P}^{\bR*}_{\nu m}(\bk) \,\Delta\Sigma^{\bR,\rm imp}_{mm'}(i\omega_{\rm n})
\,\bar{P}^{\bR}_{m'\nu'}(\bk)\;,\label{eq:sig_limband}
\end{eqnarray}
with $\bar{P}$ denoting the normalized projection between Bloch space and correlated 
subspace~\cite{ama08}.  Note that eq. (\ref{eq:g_limband}) is necessary to define the
notion of a local Green's function (eq. (\ref{eq:lgreen})) for the DMFT problem.
As for the correlated subspace, there is a choice for the 
range ${\cal W}$ of included Kohn-Sham bands in the downfolding. The object 
$\Delta\Sigma^{\rm bloch}_{\nu\nu'}$ describes the $\bk$-dependent self-energy in Bloch 
space after a double-counting correction. The latter takes care of the fact that some 
correlations are already handled on the DFT level. 
In the upfolding operation, the charge density $n(\br)$ is then decorated with DMFT
correlations, i.e. 
\begin{equation}
n(\br)=\sum \limits_{\bk,\nu\nu'}
\langle \br \vert \Psi_{\bk \nu} \rangle
\Bigl(f(\tilde{\epsilon}_{\bk \nu})\delta_{\nu \nu'}+
\Delta N_{\nu \nu'}(\bk)\Bigr)
\langle \Psi_{\bk \nu'} \vert \br \rangle\quad,
\label{eq:rho}
\end{equation}
where $\Psi$ denotes Kohn-Sham states, $f$ marks the associated Fermi function and 
$\Delta N$ refers to the DMFT Bloch-density term~\cite{lec06,ama08}. A pure band picture 
is not adequate for a many-body system and real-space excitations also matter.
Therefore, additional off-diagonal terms in the band index contribute in the 
correlated regime. This extended charge-density formulation accordingly then defines 
a new Kohn-Sham potential. Let us finally iterate on the fact that this realistic 
many-body scheme works, at heart, at finite temperature $T$. Electron states are 
subject to the full thermal impact, beyond sole occupational Fermi-function modification. 
For more formal accounts on the DFT+DMFT scheme, we refer to~\cite{geo04,kot06}.

In various problems of multi-atom unit cells, the correlated subspace encircles not only 
a single lattice site. For symmetry-equivalent sites, the self-energy is determined for 
a representative site and transferred to the remaining sites via proper symmetry 
relations. An impurity problem is defined for each symmetry-inequivalent 
site $j$ through~\cite{pot99}
\begin{equation}
{\mathcal G}^{(j)}_0(i\omega_{\rm n})^{-1}=G^{(j)}(i\omega_{\rm n})^{-1}+
\Sigma^{(j)}_{\rm imp}(i\omega_{\rm n})\quad,
\end{equation}
and the coupling is realized via the DFT+DMFT self-consistency condition invoking the 
computation of the complete lattice Green's function.

This concludes the brief sketch of the charge self-consistent DFT+DMFT 
methodology~\cite{sav01,pou07,gri12}. In the next subsection, we turn to the actually 
chosen representation to tackle delafossites.

\subsection*{Concrete setting for delafossites}
Charge self-consistent DFT+DMFT is employed for all electronic structure problems 
discussed and addressed in this overview for ${\cal A}{\cal B}$O$_2$-based delafossite. 
A mixed-basis pseudopotential method~\cite{els90,lec02,mbpp_code}, based on 
norm-conserving pseudopotentials and a combined basis of localized functions and 
plane waves is used for the DFT part. The generalized-gradient approximation in form 
of the PBE functional~\cite{per96} is utilized for the exchange-correlation functional. 
Within the mixed basis, localized functions for the transition-metal $3d$ and $4d$
shells, as well as for O($2s$,$2p$) are used to reduce the plane-wave energy 
cutoff. The latter is set to 20\,Ry for the bulk systems, and to 16\,Ry for the
heterostructures. The $k$-point mesh amounts to 13$\times$13$\times$13
for the bulk and 11$\times$11$\times$3 (9$\times$9$\times$5) for out-of-plane(in-plane)
heterostructures, respectively.
For all systems, the experimental lattice parameters are adopted and the 
internal degree of freedom $z$, governing the oxygen distance to the ${\cal A}$ plane,
is obtained from DFT structural optimization (see Tab.~\ref{tab:crystal}).
\begin{table}[b]
\begin{tabular}{C{2cm}|C{2cm} C{2cm} C{2cm}}
compound & $a$\; in \AA\ & $c$\; in \AA\ & $z$ \\ \hline
PdCoO$_2$ & 2.830 &  17.743  &  0.1132\\[0.1cm]
PdCrO$_2$ & 2.930 &  18.087  &  0.1101\\[0.1cm]
AgCrO$_2$ & 2.985 &  18.510  &  0.1095\\[0.1cm]
\end{tabular}
\caption{Experimental lattice parameters~\cite{sha71-1,sha71-2,sha71-3,ouy08} of the studied 
bulk delafossites, as well as the DFT optimized internal $z$ degree of freedom.}
\label{tab:crystal}
\end{table}

We choose the correlated subspace to be build up from the five effective 
${\cal B}$-site Wannier-like $3d$ functions as obtained from the projected-local-orbital 
formalism~\cite{ama08,ani05}, using as projection functions the linear combinations 
of atomic $3d$ orbitals which diagonalize the ${\cal B}$-site $3d$ orbital-density 
matrix. As it will be seen later, the most-relevant correlated states in delafossites are
of threefold $t_{2g}$ kind. However due to the subtle hybridization between two 
different TM sites and our further designing perspective, we stick to the more general
full $3d=\{t_{2g},e_g\}$ fivefold throughout the presented results. For selected aspects, 
a reduction of the correlated subspace to the $t_{2g}$ sector might still be an acceptable 
approximation.

A five-orbital Slater-Kanamori Hubbard Hamiltonian, i.e. including 
density-density, spin-flip and pair-hopping terms, is utilized in the correlated 
subspace, parametrized by a Hubbard $U$ and a Hund exchange $J_{\rm H}$. It reads for
orbitals $m,m'$
\begin{eqnarray}
H_{\rm int}=&&\,U\sum_{m} n_{m\uparrow}n_{m\downarrow}
+\frac 12 \sum \limits _{m \ne m',\sigma}
\hspace*{-0.2cm}\Big\{\left(U-2J_{\rm H}\right)\, n_{m \sigma} n_{m' \bar \sigma}+\nonumber\\
&&+ \left(U-3J_{\rm H}\right) \,n_{m \sigma}n_{m' \sigma}+\nonumber\\
&&\left.+J_{\rm H}\left(c^\dagger_{m \sigma} c^\dagger_{m' \bar\sigma}
c^{\hfill}_{m \bar \sigma} c^{\hfill}_{m' \sigma}
+c^\dagger_{m \sigma} c^\dagger_{m \bar \sigma}
c^{\hfill}_{m' \bar \sigma} c^{\hfill}_{m' \sigma}\right)\right\}\;,
\end{eqnarray}
with $c^{(\dagger)}_{m\sigma}$ as the annihilation(creation) operator for spin flavor
$\sigma=\uparrow,\downarrow$, and $n=c^\dagger c$. Our 
${\cal B}$ site will here be either of Co or Cr type and a value of $J_{\rm H}=0.7$\,eV 
is proper for TM oxides of that kind. Hubbard $U$ values between 3$-$4\,eV will
be chosen according to adequate onsite Coulomb integrals for similiar types of
oxides~\cite{kor98,lec09}. Note that no further Hubbard interactions are assigned
to the ${\cal A}$ site. The $d$ orbitals on those sites will be here of $4d$ kind, of formal 
$d^9$ filling and only weakly hybridizing with oxygen. Thus by any means, Coulomb 
interactions are expected much smaller than on the ${\cal B}$ site. Spin-orbit coupling
is neglected in the crystal calculations.

The encountered DMFT impurity problems in the examined delafossite materials are
solved by the continuous-time quantum Monte Carlo scheme of hybridization-expansion 
form~\cite{rub05,wer06} as implemented in the TRIQS package~\cite{par15,set16}. 
A double-counting correction of fully-localized-limit type~\cite{ani93}, utilizing 
iterated TM$(3d)$ occuptations, is applied in all calculations. 
To obtain the spectral information, analytical continuation from Matsubara space via 
the maximum-entropy method~\cite{jar96} as well as the Pad{\'e} method~\cite{vid77} 
is performed. Though the ${\cal A}$CrO$_2$ delafossites order antiferromagnetically at 
low temperatures within the CrO$_2$ layers, our investigations remain at still higher
temperatures and assume paramagnetism for all studied cases. If not otherwise stated,
the system temperature is set to $T=290$\,K.

\section*{Basic properties of C\MakeLowercase{o}O$_2$-based 
and C\MakeLowercase{r}O$_2$-based delafossites}
\subsection*{General considerations}
The delafossites PdCoO$_2$, PtCoO$_2$ and PdCrO$_2$ are metals with surprisingly high 
conductivity (see e.g.~Refs.~\cite{mac17,dao17} for recent reviews). With an 
in-plane resistivity of 2.6 $\mu\Omega$cm~\cite{hic12}, the PdCoO$_2$ compound is
designated as the most-conductive oxide at room temperature. It apparently shows
hydrodynamic flow of electrons~\cite{mol16}. Although obviously also a correlation 
effect~\cite{and11}, this feature will not be directly addressed in the present text,
but further information can be found in Refs.~\cite{gur63,sca17,var20} and references 
therein. The AgCrO$_2$ delafossite is an insulator with a charge gap of 
$\Delta=1.68$\,eV~\cite{ouy08}. While no ordering transition takes place in the Co 
compounds down to lowest temperatures, the Cr compounds display magnetic transitions into 
an antiferromagnetic (AFM) 120$^\circ$ phase below the N{\'e}el temperatures 37.5\,K 
(PdCrO$_2$)~\cite{mek95} and 21\,K (AgCrO$_2$)~\cite{ooh94,sek08}. 

From symmetry, the local electronic $d$-shell states on ${\cal A}$ and ${\cal B}$ 
sites show a trigonal splitting into $t_{2g}=\{a_{1g},e_g'\}$ and $e_g$ classes
(see also Ref.~\cite{eye08}).
The $e_g'$ and $e_g$ states are doubly degenerate, respectively. The symmetry-adapted
orbitals $|m_{\cal A,B}\rangle$ may be expressed as linear combinations of the atomic 
$d$ orbitals. Note that we choose the $x$,$y$-axis parallel to and the $z$-axis
perpendicular to the delafossite layers. On the ${\cal A}$ site, there is a 
one-to-one matching between crystal-field orbitals and atomic orbitals, i.e. 
$\ket{a_{1g}}=\ket{d_{z^2}}$, 
$\ket{e_g'(1,2)}=\ket{d_{xz},d_{yz}}$,
$\ket{e_g(1,2)}=\ket{d_{xy},d_{x^2-y^2}}$. On the ${\cal B}$ site, the identification
is based on the usual trigonal representation, reading
\begin{equation}
\left(\begin{array}{l} \ket{a_{1g}} \\ \ket{e_g'(1)} \\ \ket{e_g'(2)} 
\\ \ket{e_g(1)} \\ \ket{e_g(2)} \end{array} \right)=
\left(\begin{array}{rrrrr}
 1 &  0 & 0 & 0 & 0 \\
 0 &  a & 0 & b & 0 \\
 0 &  0 & -a & 0 & b \\
 0 &  0 & b & 0 & a \\
 0 &  b & 0 & -a & 0 \\
\end{array} \right)
\left(\begin{array}{l} \ket{d_{z^2}} \\ \ket{d_{xz}} \\ \ket{d_{yz}} \\ 
\ket{d_{xy}} \\ \ket{d_{x^2-y^2}}
\end{array} \right)\;.\label{eq:ltotraf}
\end{equation}
The values $a,b$ may be obtained from diagonalizing the DFT orbital density matrix
for the respective $3d$ shell after convergence of the crystal calculation. The nominal 
${\cal B}^{3+}$ ion is usually in a low-spin configuration with its $e_g$ states
higher in energy and mostly empty. The collected DFT crystal-field levels of the $d$ 
states on the respective ${\cal A}$ and ${\cal B}$ sites, along with the $a,b$ 
coeffcients, are given in Tab.~\ref{tab:cf}

Spin-orbit effects are assumed not to play a decisive role for the transport properties, 
however, they might have some influence in the magnetically ordered phases. The Co ion with
configuration Co$^{3+}(3d^6)$ has a closed $t_{2g}$ subshell in the local limit,
which explains the absence of magnetic ordering in (Pd,Pt)CoO$_2$. On the other hand, 
Cr$^{3+}(3d^3)$ has a half-filled $t_{2g}$ subshell in that limit. Therefore,
correlation effects, which should predominantly originate from the TM($3d$) ions,
are naturally expected stronger for (Pd,Ag)CrO$_2$. On the ${\cal A}$ sites, formally,
the Pd ions have a $4d^9$ and the Ag ions a $4d^{10}$ oxidation state. Because of 
the filled Ag($4d$) shell and the insulating nature, the Cr electrons in AgCrO$_2$ are 
localized in a Mott fashion. Hence, AgCrO$_2$ is a combined 'band-Mott' insulator.

In view of some general Mott-relevant classifications given in the Introduction, 
one can state that the Co,Cr$(3d)$-derived states are of multiorbital type, at or 
very close to half filling and expectedly located in a Mott-Hubbard rather than 
charge-transfer regime. Significant orbital-selective or Hund-metal tendencies are also 
not expected because of weak orbital differentiation and the half-filled nature.
In the following, we will first discuss the more detailed electronic structure of
PdCoO$_2$, PdCrO$_2$ and AgCrO$_2$ from a nonmagnetic Kohn-Sham DFT viewpoint. Previous
DFT accounts of these system may be found e.g. in 
Ref.~\cite{ses98,eye08,ong10,ong12,sob13,noh14,gru15,bil15}.
\begin{table}[t]
\begin{tabular}{L{1.5cm}|R{1cm} R{1cm} R{1cm} R{1.25cm} R{1cm}}
compound & $\varepsilon_{a_{1g}}^{\cal A,B}$ & $\varepsilon_{e_g'}^{\cal A,B}$ & 
         $\varepsilon_{e_g}^{\cal A,B}$ & a & b \\ \hline
               & -1082 &  -2177  & -1953 & &  \\
\bw{PdCoO$_2$} & -1320 &  -1415  &  -465 & 0.621 & 0.784\\[0.1cm]
               &  -969 &  -2057  & -1847 & &  \\
\bw{PdCrO$_2$} &  -336 &  -493   &   831 & 0.586 & 0.810\\[0.1cm]
               & -3195 &  -4139  & -4080 & & \\ 
\bw{AgCrO$_2$} &  -249 &   -459  &   743 & 0.548 & 0.837\\[0.1cm]
\end{tabular}
\caption{DFT crystal-field levels on ${\cal A}$ (first row) and ${\cal B}$ (second row) 
site in the investigated delafossites, as well as orbital coefficients $a,b$ on the 
${\cal B}$ site. All energies in meV.}
\label{tab:cf}
\end{table}
\begin{figure*}[t]
\begin{center}
\includegraphics*[width=17.75cm]{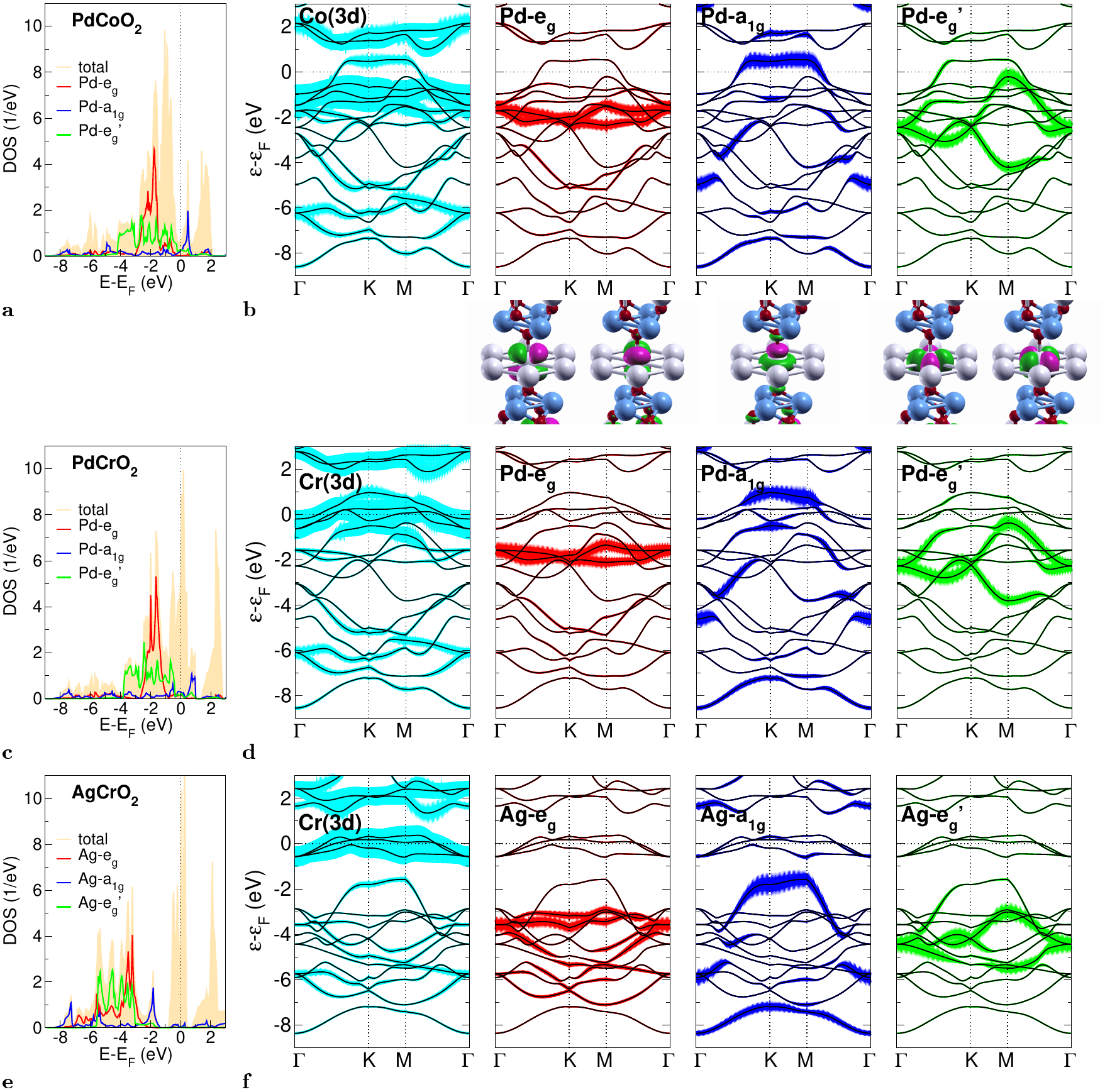}
\end{center}
\caption{
{\bf DFT electronic structure of PdCoO$_2$, PdCrO$_2$ and AgCrO$_2$}. 
({\bf a}) Total and orbital-resolved ${\cal A}$-site$(4d)$ density of states. 
({\bf b}) Band structure along high symmetry lines in the $k_z=0$ plane, with 
fatbands marking the ${\cal B}$-site$(3d)$ and the orbital-resolved 
${\cal A}$-site$(4d)$ character.
Additonally, the corresponding real-space Pd$(4d)$ projected local orbitals are
provided for the case of PdCoO$_2$: Pd (grey), Cr (lightblue) and O (red).
({\bf c},{\bf d}) Same as ({\bf a},{\bf b}) but for PdCrO$_2$. 
({\bf e},{\bf f}) Same as ({\bf a},{\bf b}) but for AgCrO$_2$.
\label{fig:dft}}
\end{figure*}

\subsection*{DFT picture}
For PdCoO$_2$, Figs.~\ref{fig:dft}a,b display the spectral DFT properties, 
namely density of states (DOS) and band structure of this metallic delafossite,
as well as provide plots of the Wannier-like Pd$(4d)$ orbitals. 
As expected, the Pd$(4d)$ states are largely occupied with a bandwidth ($W$) hierachy of 
$W_{a_{1g}}>W_{e_g'}>W_{e_g}$. As shown in Fig.~\ref{fig:dft}b, the Co$(3d)$ weight
is mostly located in the bands close to and above the Fermi level 
$\varepsilon_{\rm F}^{\hfill}$, with a single band crossing $\varepsilon_{\rm F}^{\hfill}$. 
The latter dispersion, which we denote in the 
following 'cPd', is dominantly of mixed Pd$(4d)$ and partial Co$(3d)$ kind. 
In more detail from the Pd site, Pd-$a_{1g}$ and
Pd-$e_g'$ have a comparable weight on the $\varepsilon_{\rm F}^{\hfill}$-crossing
regime of that most-relevant band. Note that the $e_g'$ orbitals are the ones with 
the strongest in-plane character (see Fig.~\ref{fig:dft}b). As a further note,
though the band-filling Co$(3d)$ character resembles the original Co$(3d^6)$ picture,
from the hybridizations at the Fermi level a completely inert Co-$t_{2g}$ subshell is 
not truly justified. The DFT fermiology and dispersions at low energy are in good
agreement with data from angle-resolved photoemission spectroscopy (ARPES) 
measurements~\cite{noh09} and de Haas-van Alphen studies~\cite{hic12}. Thus plain
DFT seemingly provides already an adequate description of key PdCoO$_2$ features.

The spectral DFT properties of PdCrO$_2$ are shown in Figs.~\ref{fig:dft}c,d. Contrary to 
PdCoO$_2$, the ${\cal B}$-site states of Cr$(3d)$ character are much less filled, and the 
three $t_{2g}$-dominated bands are right at the Fermi level. On the other hand, the 
Pd$(4d)$ character at $\varepsilon_{\rm F}^{\hfill}$ is minor. This low-energy picture 
of the dispersions however strongly disagrees with available experimental data from
ARPES~\cite{sob13,noh14} and quantum oscillations~\cite{ok13,hic15}. In experiment,
there is also only a single band crossing the Fermi level, quite similarly as in 
PdCoO$_2$. This discrepancy is due to the neglect of strong electronic correlations
in conventional DFT, which misses the Mott-localized character of the CrO$_2$ layers.
Partial agreement with experiment concerning the dispersions can be achieved within 
spin-polarized DFT~\cite{ong12,sob13,noh14,bil15}, accounting also for the magnetic
ordering at low temperatures. But this Slater-type handling of the Cr$(3d)$ states
is not truly describing the underlying physics correctly. For instance, the single cPd
dispersion holds for temperatures well above the magnetic-ordering 
temperature~\cite{sob13}, therefore the gapping of Cr$(3d)$ is not linked to ordered 
magnetism.

Finally, the AgCrO$_2$ compound would be insulating in DFT if the Cr-$t_{2g}$ states were 
not located again at the Fermi level (see Figs.~\ref{fig:dft}e,f). The Ag$(4d)$ states
are filled and would give rise to a band insulator. The missing correlation effects on
Cr become most evident in this delafossite. Note the prominent Ag-$a_{1g}$ dominated
band just below the Cr-$t_{2g}$ bands in energy and with a nearly flat dispersion 
along K-M. It bears striking resemblance to the former low-energy cPd band in 
PdCoO$_2$. In fact as we will see in the following, this present band will just form
the highest valence band in true AgCrO$_2$ once correlations are properly included.
Furthermore for the same reason, the akin band in PdCrO$_2$ (yet there with stronger
Pd-$e_g'$ character) will be shifted to $\varepsilon_{\rm F}^{\hfill}$, giving rise to 
the experimentally revealed single-sheet fermiology.

\subsection*{DFT+DMFT picture}
Let us now turn to the an improved description of the given delafossites, arising from 
the inclusion of correlation effects within DFT+DMFT. Concerning the Hubbard interaction, 
the value $U=3$\,eV is assigned to the $3d$ states of Co and Cr in PdCoO$_2$ and 
PdCrO$_2$, respectively.
For Cr$(3d)$ in AgCrO$_2$, the somewhat larger value of $U=4$\,eV is used in order to
comply with the weaker screening because of the (nearly) filled Ag$(4d)$ shell. A larger
part of this subsection builds up on results and discussions provided in 
Refs.~\cite{lec18,lec20}.\\

\begin{figure*}[t]
\begin{center}
\includegraphics*[width=16cm]{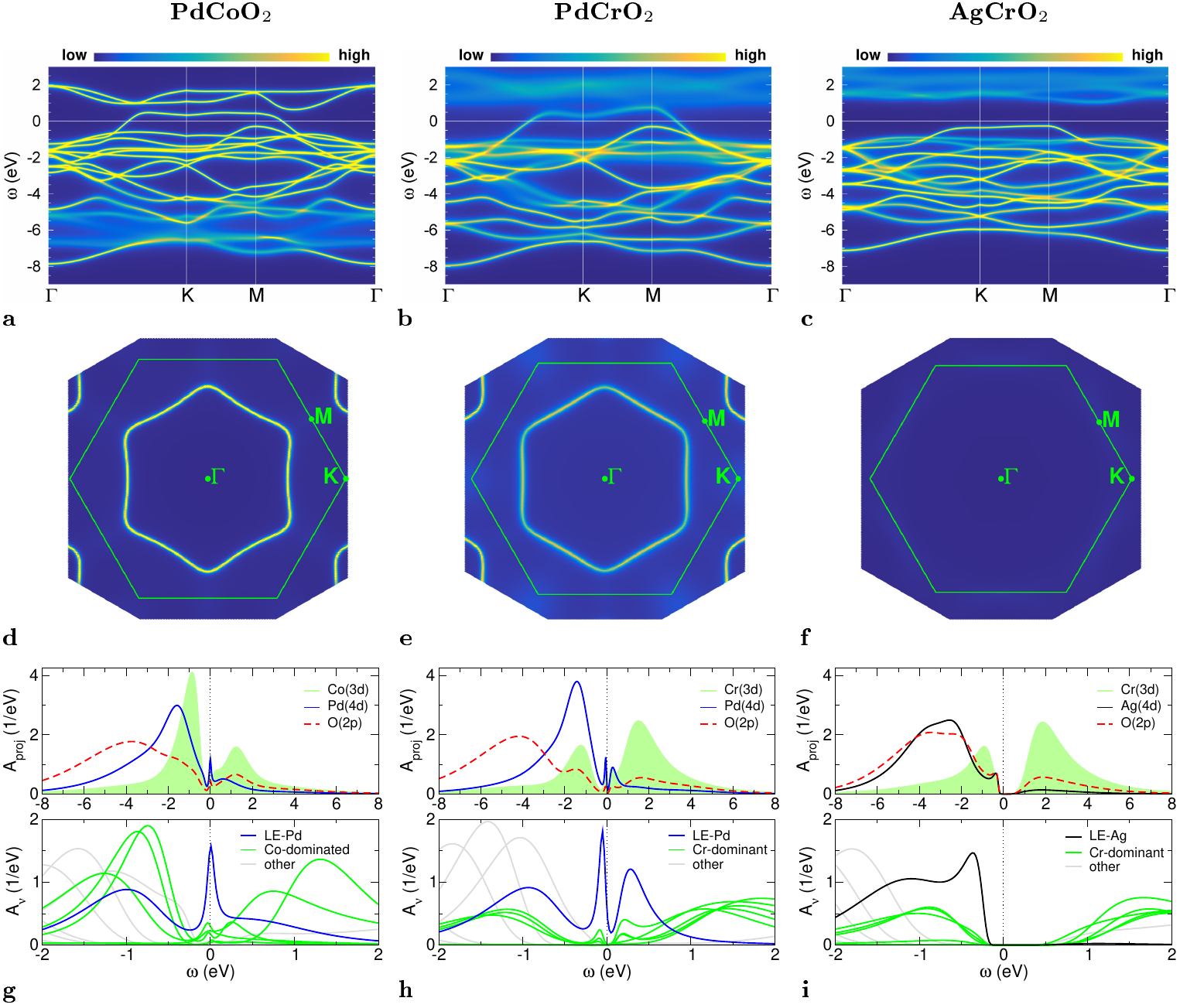}
\end{center}
\caption{
{\bf Spectral properties of PdCoO$_2$, PdCrO$_2$ and AgCrO$_2$ according to paramagnetic
DFT+DMFT}. ({\bf a}-{\bf c}) {\bf k}-resolved spectral function $A({\bf k},\omega)$ 
along high symmetry lines in the $k_z=0$ plane.
({\bf d}-{\bf f}) Interacting Fermi surface in the $k_z=0$ plane.
({\bf g}-{\bf i}) Site- and orbital-projected spectral function $A_{\rm proj}(\omega)$ (top) and 
Bloch-resolved spectral function $A_{\nu}(\omega)$, see text (bottom), respectively.
\label{fig:bmft}}
\end{figure*}
Figure~\ref{fig:bmft} exhibits the spectral summary for the three compounds. In the case
of the Co compound, the changes compared to DFT appear minor, as already expected from
the simplest picture of a closed Co-$t_{2g}$ subshell. The dispersions, which now describe 
true quasiparticle features, are hardly modified at lower energy. Quite on the contrary, 
the QP dispersion for PdCrO$_2$ has changed dramatically (see Fig.~\ref{fig:bmft}b); the 
DFT-original Cr bands at $\varepsilon_{\rm F}^{\hfill}$ have disappeared and instead, a single 
cPd dispersion as in PdCoO$_2$ crosses the Fermi level. This result brings theory eventually 
in line with experimental findings~\cite{sob13,noh14,ok13,hic15}. Also for AgCrO$_2$, the 
DFT+DMFT approach settles the comparison with experiment, namely by identifying the 
insulating nature with a compatible gap of $\sim 1.8$\,eV.
While the latter delafossite shows of course no Fermi surface, the fermiology of 
PdCoO$_2$ and PdCrO$_2$ in Figs.~\ref{fig:bmft}d,e becomes rather similar with 
interactions. A single-sheet interacting Fermi surface, comprising a single electron, 
is centered around $\Gamma$ and has a hexagonal shape with some warping. Note that
this warping is somewhat stronger in the case of the Co compound.

Two functions are provided to discuss the ${\bf k}$-integrated spectra (see 
Fig.~\ref{fig:bmft}h-j). First, the site- and orbital projected spectral function
$A_{\rm proj}(\omega)$, defined by projecting the Bloch-resolved spectral function
$A_{\nu}(\bk,\omega)$ with Bloch index $\nu$ onto a chosen site-orbital and summing
over $\nu, \bk$. Note that this function is comparable but strictly not identical to the 
local spectral function $A_{\rm loc}$, which is obtained from analytical continuation 
of the local Green's function. Second, it proves instructive to also plot directly
$A_{\nu}(\omega)$, i.e. the $\bk$-integrated Bloch-resolved spectrum. This allows us 
to trace the behavior of the former DFT bands upon interaction and displays the QP
formation originating in Bloch space. 

The projected spectrum of PdCoO$_2$ exhibits the near subshell filling of Co$(3d)$ and 
the Pd dominance of the low-energy QP peak at the Fermi level. While on a first glance, 
$A_{\rm proj}$ for the Cr$(3d)$ spectrum in PdCrO$_2$ looks similar to the previous 
Co$(3d)$ one, the physics is completely different; the Cr-$t_{2g}$ states are in a Mott 
state and therefore their spectral weight is shifted to deeper energies up and below the 
Fermi level, i.e. to upper and lower Hubbard bands. The Cr-$e_g$ states are mostly empty, 
but show also strong incoherence effects in Fig.~\ref{fig:bmft}b. Mott criticality in the 
CrO$_2$ layers has been originally suggested by several experiments from strong hints for 
localized Cr$^{3+}$ $S=3/2$ spins~\cite{tak09,noh14,hic15}. The QP peak at low energy is of 
dominant Pd$(4d)$ character, therefore confirming the previously announced mechanism 
of a correlation-induced shift of a DFT-original deeper lying Pd-dominated band towards 
$\varepsilon_{\rm F}^{\hfill}$. The projected AgCrO$_2$ spectrum shows again the
Mott-insulating Cr$(3d)$ part along with the band-insulating Ag$(4d)$ part. The plots of 
$A_{\nu}(\omega)$ render obvious that for PdCoO$_2$ the low-energy QP is for the
most part constituted from a single Pd-dominated Bloch dispersion, which we call 
LE-Pd. The same holds for the PdCrO$_2$ case. Yet importantly, both QP peaks display
the hybridizing contribution of Co/Cr-dominated functions, and the LE-Pd function
moreover exhibts significant energy dependence. Both features point to the relevance
of the subtle impact of electronic correlation onto the low-energy regime. Or in other
words, the 'single-band' dispersion crossing $\varepsilon_{\rm F}^{\hfill}$, though not
dominated by the strongly-interacting ${\cal B}$-site $3d$ orbitals, still carries
subtle effects of correlations which most certainly rule (parts of) the challenging
delafossite physics. But be aware of our difference in nomenclature; 'cPd' denotes the 
complete single low-energy dispersion, while 'LE-Pd' marks the most dominant $A_\nu$
contribution to it. The corresponding $A_{\nu}$ plot for AgCrO$_2$ shows that the
band-insulating character part is equally dominated by a single LE-Ag dispersion,
with the Mott-insulating character part once again carried by Cr$(3d)$.

Finally note that the O$(2p)$ states are mostly aligned with Ag$(4d)$ in AgCrO$_2$,
whereas they are located significantly deeper in energy than Pd$(4d)$ in PdCoO$_2$ 
and PdCrO$_2$. Since there are no strong charge-transfer effects expected from 
$t_{2g}$-based Co,Cr$(3d)$, the present treatment of excluding explicit Coulomb 
interactions within O$(2p)$ should be reliable. There might be however scenarios, e.g.
the hole doping of AgCrO$_2$, where O$(2p)$-based interaction effects could become 
non-negligible in delafossites. Extensions of conventional DFT+DMFT, e.g. as described
in Ref.~\cite{lec19}, are available to treat such effects.\\

\begin{figure}[t]
\begin{center}
\includegraphics*[width=8.25cm]{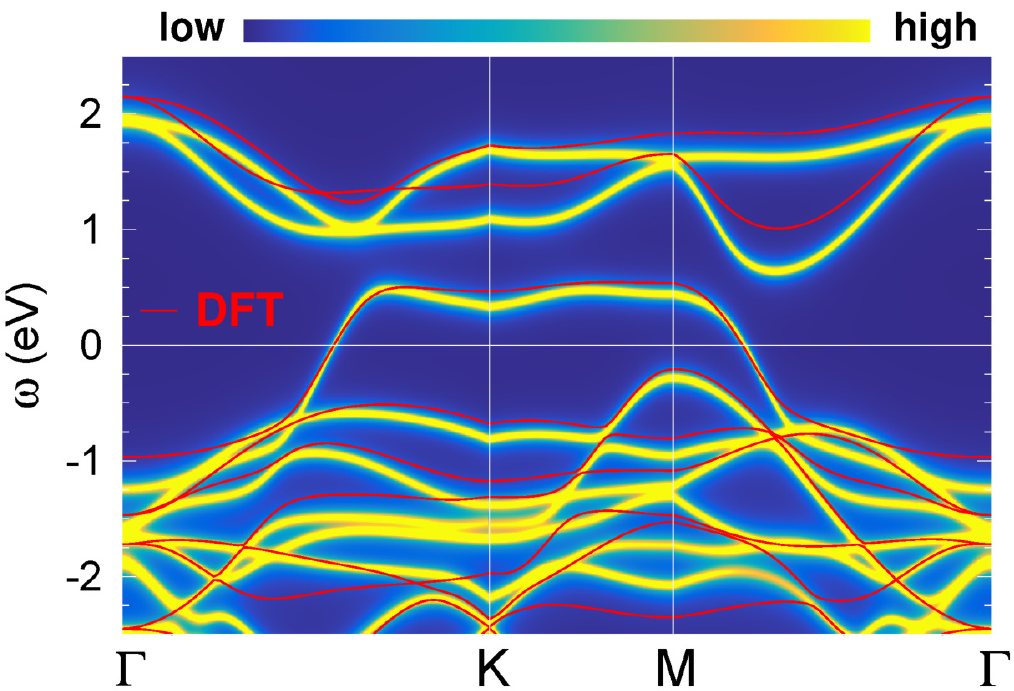}
\end{center}
\caption{
{\bf {\bf k}-resolved spectral function of PdCoO$_2$}. Comparison between the DFT band structure
(red) and the DFT+DMFT result. 
\label{fig:detpdco}}
\end{figure}

After this overview, we want in the following discuss and comment on relevant underlying 
electron correlation aspects. First, the atomic-like picture
of a fully-closed Co-$t_{2g}$ subshell in the PdCoO$_2$ compound is of course an 
idealization. The DFT+DMFT occupancies per single orbital on Co at room temperature 
amount to $\{n_{e_g'},n_{a_{1g}}\}=\{1.90,1.93\}$, thus there is still about 4\% local
$t_{2g}$ doping. The associated charge fluctuations together with the hybridizations on 
the low-energy QP dispersion point to a subtle connection between the Pd layer and the
CoO$_2$ layer. This is underlined by the spectral comparison shown in 
Fig.~\ref{fig:detpdco}, where one may observe that the 
$\varepsilon_{\rm F}^{\hfill}$-slope of the cPd dispersion, which is proportional to 
the Fermi velocity,
has slightly increased with correlations. Usually, strong local electronic correlations 
lead to a reduction of the QP Fermi velocity, associated with a bandwidth renormalization 
toward smaller values. Yet here, the explicit Coulomb interactions are active in the 
CoO$_2$ layer and the cPd dispersion with dominant Pd$(4d)$ weight mainly originates from 
the 'non-interacting' Pd layer. Seemingly, an implicit nonlocal effect of correlation 
steepens the dispersion, at least on a proof-of-principle level. This could be a 
contributing factor to the high conductivity of PdCoO$_2$. Again, (local) correlation 
effects appear not decisive in PdCoO$_2$, but their role still deserves further exploration.
\begin{figure*}[t]
\begin{center}
\includegraphics*[width=16.5cm]{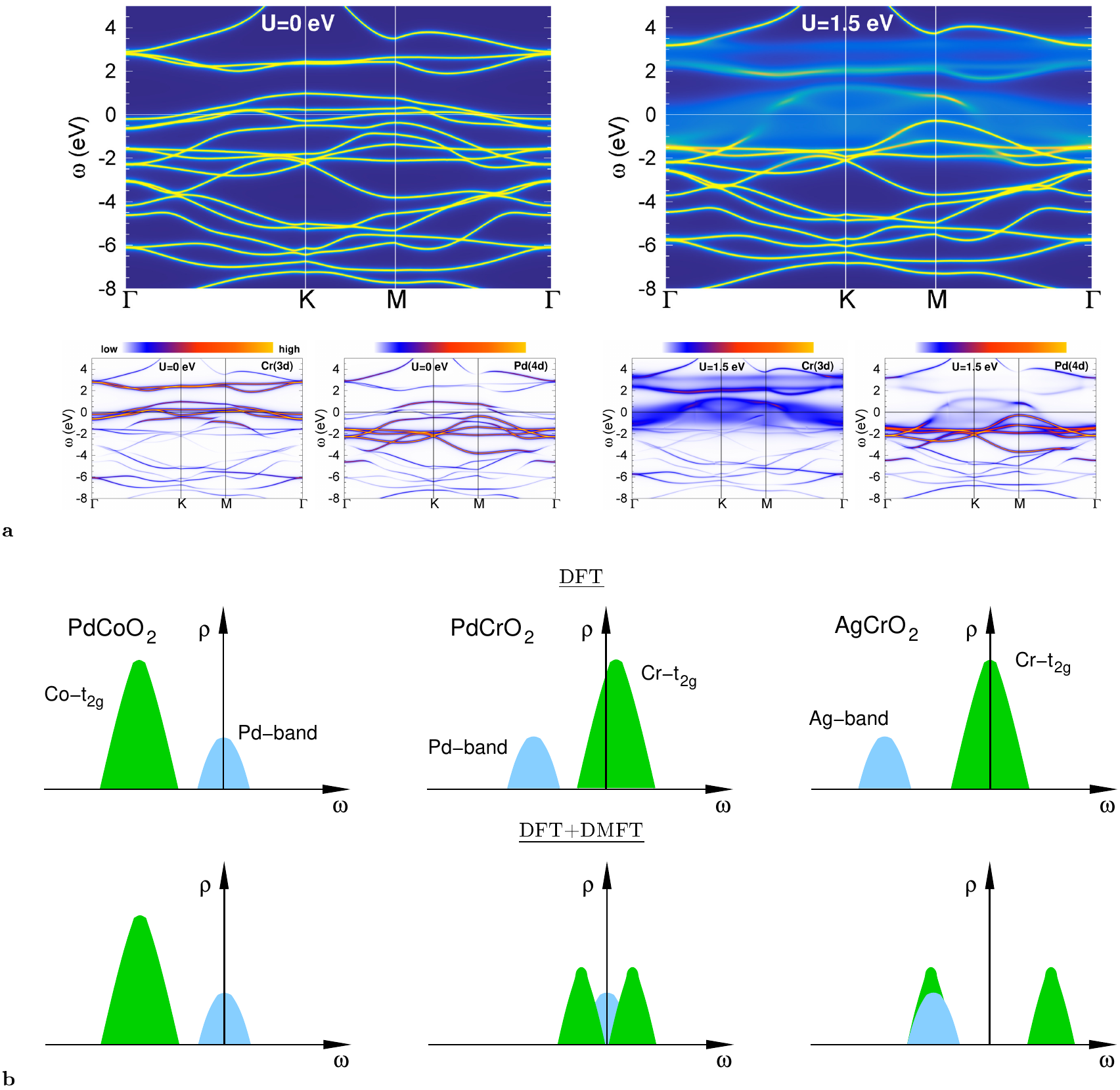}
\end{center}
\caption{{\bf On the correlation features in delafossites}. 
({\bf a}) Spectral properties of PdCrO$_2$ for different interaction strengths. 
Left panel: $U=0$\,eV, i.e. DFT bands. Right panel: $U=1.5$\,eV. 
Top: spectral function, bottom: fatbands for Cr$(3d)$ (left) and Pd$(4d)$ (right) 
(from~\cite{lec20}).  
({\bf b}) Sketch of the basic differences in the {\bk}-integrated spectral function 
$\rho(\omega)$ for the most-relevant four bands close to the Fermi level among the studied 
delafossites. Top: DFT, bottom: DFT+DMFT. 
Left: PdCoO$_2$: Co-$t_{2g}$ bands occupied, uppermost Pd$(4d)$-band half filled;
middle: hidden-Mott PdCrO$_2$: Cr-$t_{2g}$ bands 1/3 filled, uppermost Pd$(4d)$-band 
occupied;
right: band-Mott insulating AgCrO$_2$: Cr-$t_{2g}$ bands half filled, uppermost 
Ag$(4d)$-band occupied.
\label{fig:delcorr}}
\end{figure*}

The impact of electronic correlations is obviously crucial for PdCrO$_2$. On a more formal
level, interactions lead to a metal-to-metal transition between a system with
Cr$(3d)$-dominated threefold dispersion at weak coupling and a system with 
Pd$(4d)$-dominated single dispersion at strong coupling. The question arises how this
apparent quantum phase transition takes place with increasing interaction strength $U$.
Figure~\ref{fig:delcorr}a displays the spectral function and fatbands for the DFT limit
('$U=0$') and for $U=1.5$\,eV, i.e. half the assumed correct interaction strength in 
PdCrO$_2$, in direct comparison. Of course, the $U=0$ data is identical to what is
shown in Fig.~\ref{fig:dft}d, rendering the mechanism for the transition clear:
The three bands at the Fermi level are filled with two electrons, and hence four electrons
populate the altogether four bands when counting down in energy from 
$\varepsilon_{\rm F}^{\hfill}$. These four bands are of mixed Cr$(3d)$, Pd$(4d)$ character,
with dominance from the $3d$ sector. Due to the given band entanglement, strong
correlations transform three of them into Hubbard bands, and leave a resulting one
with half filling at the Fermi level. Interestingly, for the intermediate coupling
(right panel of Fig.~\ref{fig:delcorr}a), the system adopts a 'strange' situation. 
The Cr-$t_{2g}$
dispersions are very incoherent and not yet Mott localized, and the cPd dispersion is not
yet fully established coherently. Note that especially the cPd dispersion, which appears
weakly-interacting at strong and weak coupling, is intriguingly affected by correlations 
close to the given metal-to-metal transition. This underlines the intricate inter-layer
physics that is at work in PdCrO$_2$ with its 'hidden-Mott' state. Note that in a pure
model context, at least three bands would be necessary to capture a hidden-Mott scenario:
one fully filled band and two bands with overall quarter filling when putting
interactions to zero. Strong interactions should then transfer one electron from the
filled band to both remaining bands, rendering them half filled.

From a model perspective of correlated electrons, Kondo-lattice type of Hamiltonians
describing strongly-interacting sites within a Fermi sea~\cite{mot74,don77} have been
discussed as a starting perspective for PdCrO$_2$~\cite{lec18}. Such a framework has 
then indeed put into practise in order to account for the coupling of the Cr spins to 
the Pd layer in the magnetically-ordered state~\cite{sun20}. However, as already 
mentioned in Ref.~\cite{lec18}, a standard Kondo-lattice model of spins coupled to 
free electrons appears too simplistic to cover the full complexity of the above 
described hidden-Mott physics~\cite{kom20}. Modelling the electronic correlations that 
originate from the CrO$_2$ layer and spanning over to the Pd layer in a comprehensive way 
has most definitely to account for the outlined metal-to-metal transition.

Because of the $4d^{10}$ state of silver in AgCrO$_2$, an intricate band entanglement 
as in PdCrO$_2$ is missing. In the DFT limit, the three Cr-$t_{2g}$ bands at the Fermi 
level are already half filled with three electrons. Thus the internal Mott transition 
in the CrO$_2$ layers does not lead to a metal-to-metal transition, but to a more
ordinary metal-to-insulator transition with increasing $U$. But there is a twist; the
valence-band maximum of insulating AgCrO$_2$ is dominated by silver (and oxygen) 
character, highlighting the band-insulating aspect of the system. The compound is
therefore best coined as band-Mott insulator. 

To emphasize the key differences of the given delafossites from a minimal perspective, 
Fig.~\ref{fig:delcorr}b summarizes the main features from the noted four-band perspective 
of ${\cal B}$-site derived $t_{2g}$ bands and ${\cal A}$-site derived uppermost $4d$ 
band part.

\section*{Mott design of correlated delafossites}
\subsection*{General considerations}
We have seen in the previous section that in an interacting many-body sense, the revealed
correlation effects in PdCoO$_2$ and AgCrO$_2$ are apparently not yet of particular 
breathtaking kind. The former compound is a straightforward metal with, from the current 
viewpoint, weak impact of correlations. 
The latter compound harbors strong correlations and is a combined band-Mott insulator, 
yet a Mott-insulating state per se is not a spectacular state of matter. On the other
hand, the PdCrO$_2$ compound seems quite exotic with its entanglement between metallic
and Mott-insulating characteristics. 
However in the equilibrium state, PdCrO$_2$ behaves like an ordinary metal, even
across the N{\'e}el temperature and in the ordered AFM phase~\cite{tak09}. The 
hidden-Mott state in the system is seemingly behaving like a 'sleeping dragon'. For
its awakening and the display of more exciting physics, one has to drive the compound
'out of its comfort zone' by disturbance and further design.

Hence in the following subsections the focus will be on possible theoretical ways how to 
'wake up the dragon' and to create new correlation phenomenology out of the intriguing
scenario found in PdCrO$_2$.
We will here only briefly comment on point-defect and pressure/strain effects in the
next subsection, and afterwards will discuss in some more detail the effects of 
heterostructuring. An overview on recent experimental 
activities towards engineering correlated delafossites will close this section.

\subsection*{Defect engineering, pressure and strain\label{deldef}}
Chemical doping either iso-valent or of charge-doping kind is a traditional route
to modify a given electronic structure. The hole- and electron doping of cuprates
by substitutional impurities which transfers a stoichiometric Mott insulator into
a high-$T_{\rm c}$ superconductor represents the most famous example~\cite{bed86}.
Melting the instrinsic Mott insulator in PdCrO$_2$ by defect-induced charge
doping becomes indeed possible from supercell DFT+DMFT calculations in a 
corresponding dense-defect regime with symmetry breakings due to point 
defects. On the other hand, minor charge doping on the Cr site performed in a 
virtual-crystal approximation leads to a transfer of doping charge from the CrO$_2$
layers to the Pd layers, thus doping mainly the metallic band. We refer the reader to 
Ref.~\cite{lec18} for further details. 

Additionally, iso-valent chemical doping may be promising from the metal-to-metal
transition viewpoint given above. Namely, introducing e.g. Mo impurities on the Cr 
sublattice should lead to a reduction of the effective $U$ in the $d$-shell on 
the ${\cal B}$ site, enabling access to the intriguing intermediate-coupling regime.

Application of pressure or strain can also result in a relevant change of the electronic
structure. For instance, uniaxial pressure/strain along the $c$-axis would 
modify the Pd and CrO$_2$ layer separation, which may effect the band entanglement 
and the hidden-Mott physics. There is the possibility of driving the discussed
metal-to-metal transition by metallizing the CrO$_2$ planes via applied pressure.

\subsection*{Out-of-plane PdCrO$_2$-AgCrO$_2$ heterostructures:
correlated semimetallic states}
As noted in the Introduction and hopefully became clearer in the previous 
sections, delafossites
may be viewed as natural heterostructures with different electronic characteristics in
the ${\cal A}$- and ${\cal B}$O$_2$-layers. It may be therefore obvious that a merging of
delafossite physics and the ever-growing field of oxide heterostructures (see e.g.
Refs.~\cite{zub11,hwa12} for reviews) could turn out as a frutiful combination. 

In fact, heterostructures from combining PdCrO$_2$ and AgCrO$_2$ may be of particular
interest. Both compounds have similar lattice parameters, resulting in a minor mismatch,
and differ only by one electron in the ${\cal A}$-site valence. However their electronic
phenomenology, i.e. hidden-Mott metal vs. band-Mott insulator, is quite different. 
Heterostructuring both delafossites provides therefore a specific doping scenario: by keeping
the local environment rather undisturbed, filling modifications in the Cr-$t_{2g}$ manifold
(i.e. green spectra in middle and right part of Figs.~\ref{fig:delcorr}b) 
inbetween the DFT values may be triggered. In the following we denote the DFT filling 
fraction of the Cr-$t_{2g}$ states by $\alpha$. Then the given heterostructures formally 
interpolate directly between the 
hidden-Mott metal ($\alpha=1/3$) and the band-Mott insulator ($\alpha=1/2$), and should show 
via which path both phases are connected. 

Alternate stackings of Pd, CrO$_2$ and Ag layers, i.e. straightforward 
out-of-plane heterostructures, are a natural realization of such a scenario.
Note that a 'simple' interface construction
between wide blocks of PdCrO$_2$ and AgCrO$_2$ might not be an ideal research object,
since presumably, most of the interface modifications to transport would be masked by 
the metallic PdCrO$_2$ block.
However, another type of heterostructure may also be promising, namely an in-plane 
variation within the ${\cal A}$ layer from intermixing the Pd and Ag content in an 
ordered fashion. This type should not be immediately linked to the doping scenario 
described above, because of the inherent change of the ${\cal A}$-layer character. 
But it could benefit from the specific delafossite layer (dis)entangling, and
possibly give rise to different scenarios of a designed 2D lattice within a Mott 
background.\\

\begin{figure}[t]
\begin{center}
\includegraphics*[width=8.5cm]{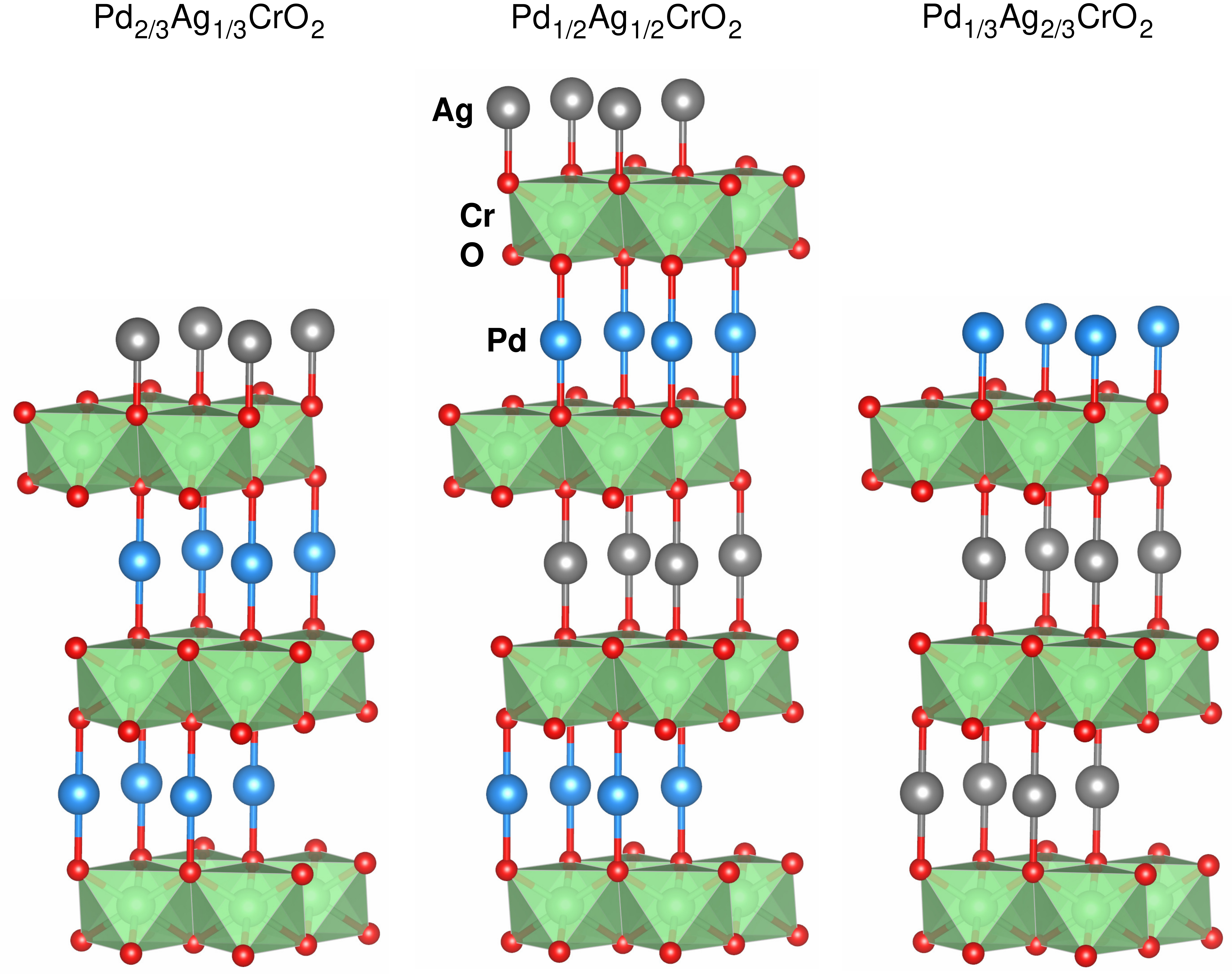}
\end{center}
\caption{{\bf Crystal structures of the designed out-of-plane heterostructures}.
Pd$_{\nicefrac{2}{3}}$Ag$_{\nicefrac{1}{3}}$CrO$_2$, 
Pd$_{\nicefrac{1}{2}}$Ag$_{\nicefrac{1}{2}}$CrO$_2$ and 
Pd$_{\nicefrac{1}{3}}$Ag$_{\nicefrac{2}{3}}$CrO$_2$
(from left to right) result from different stackings along the $c$-axis. 
Pd: blue, Ag: grey, Cr: green and O: red. From Ref.~\cite{lec20}.
\label{fig:hstruc}}
\end{figure}

Three heterostructures with different PdCrO$_2$/AgCrO$_2$ stacking along the $c$-axis
are designed, namely Pd$_{\nicefrac{2}{3}}$Ag$_{\nicefrac{1}{3}}$CrO$_2$, 
Pd$_{\nicefrac{1}{2}}$Ag$_{\nicefrac{1}{2}}$CrO$_2$ and 
Pd$_{\nicefrac{1}{3}}$Ag$_{\nicefrac{2}{3}}$CrO$_2$ 
(see Fig.~\ref{fig:hstruc}). The lattice parameters are chosen from linear-interpolating
the respective experimental data of the bulk structures (cf. Tab.~\ref{tab:crystal}),
and the Hubbard $U$ (identical on every Cr site) is also correspondingly interpolated 
from the limiting delafossite cases (see Ref.~\cite{lec20} for further details).
\begin{figure*}[t]
\begin{center}
\includegraphics*[width=16.5cm]{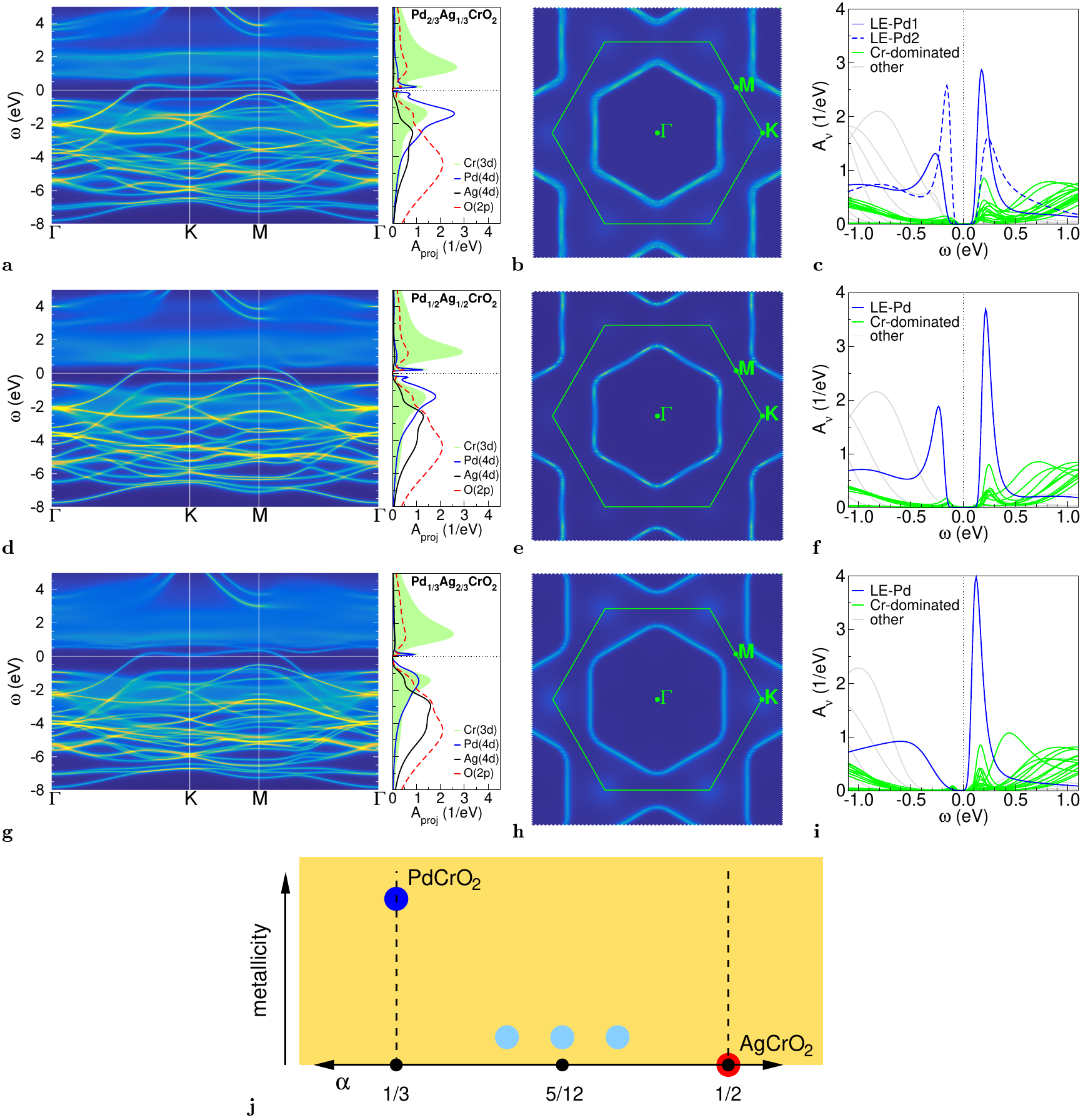}
\end{center}
\caption{{\bf Paramagnetic DFT+DMFT spectral data for the out-of-plane heterostructures}. 
({\bf a}-{\bf c}) Pd$_{\nicefrac{2}{3}}$Ag$_{\nicefrac{1}{3}}$CrO$_2$, 
({\bf d}-{\bf f}) Pd$_{\nicefrac{1}{2}}$Ag$_{\nicefrac{1}{2}}$CrO$_2$ and 
({\bf g}-{\bf i}) Pd$_{\nicefrac{1}{3}}$Ag$_{\nicefrac{2}{3}}$CrO$_2$.
({\bf a},{\bf d},{\bf g}) Spectral function $A({\bf k},\omega)$ along high-symmetry lines in the $k_z=0$
plane of reciprocal space (left) and {\bf k}-integrated site- and orbital-projected 
spectral function (right). ({\bf b},{\bf e},{\bf h}) Fermi surface for $k_z=0$ within the first 
Brillouin zone (green hexagon). (c,f,i) {\bf k}-integrated Bloch contribution 
$A_{\nu}(\omega)$ with characterization of dominance. From Ref.~\cite{lec20}.
({\bf j}) Schematic resulting room-temperature metallicity with respect to the DFT filling 
factor $\alpha$ of the Cr-$t_{2g}$ bands. 
Blue: metal PdCrO$_2$, red: band-Mott insulator AgCrO$_2$,  
and lightblue: CIS Pd$_{\nicefrac{2}{3}}$Ag$_{\nicefrac{1}{3}}$CrO$_2$, 
Pd$_{\nicefrac{1}{2}}$Ag$_{\nicefrac{1}{2}}$CrO$_2$ and 
Pd$_{\nicefrac{1}{3}}$Ag$_{\nicefrac{2}{3}}$CrO$_2$ (from left 
to right).
\label{fig:hdmft}}
\end{figure*}

Before discussing the numerical results, let us briefly brainstorm about the 
electronic structure condition. From the PdCrO$_2$ perspective, the additional
blocking layers of Ag kind as well as the stronger Mott-insulating character induced
therefrom into the CrO$_2$ layers should increase correlations within the Pd layers, too.
But this will again happen in a more subtle way than in standard correlated systems,
where an associated Coulomb repulsion for such a layer is increased when looking for
stronger correlation effects. Remember that there is no Hubbard $U$ on Pd and all
correlation increase has to take place in a nonlocal way from the surrounding layers.
Thus, the present heterostructures pose a quite original correlation problem at low 
energy; a half-filled Pd layer without intra-layer interaction, subject to rising 
'Coulomb pressure and confinement' imposed from the neighboring layers. How does the
single electron of dominant Pd$(4d)$ character cope with that situation?
\begin{figure*}[t]
\begin{center}
\includegraphics*[width=14cm]{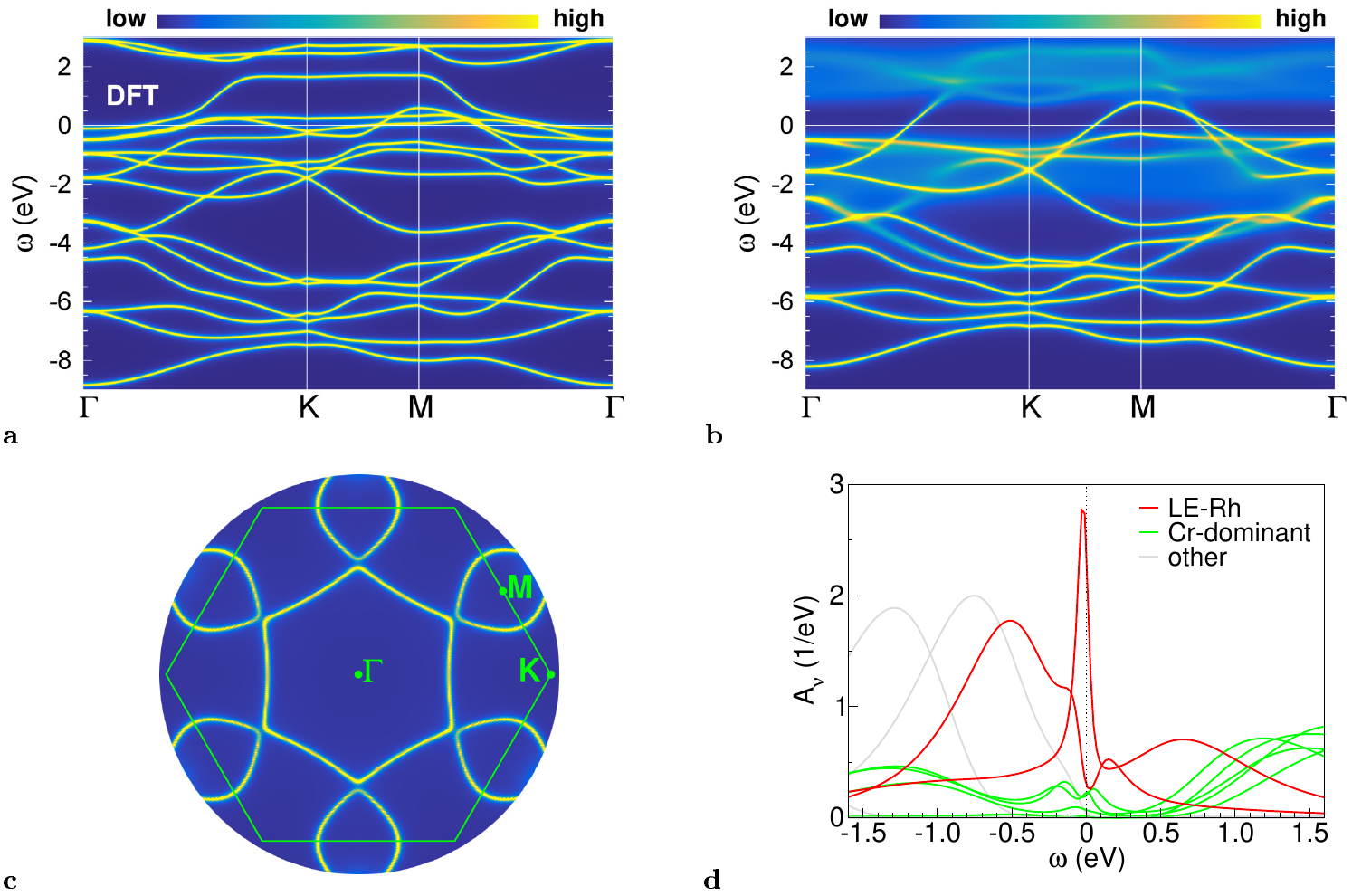}
\end{center}
\caption{{\bf Spectral properties of hypothetical RhCrO$_2$}. ({\bf a}) DFT band structure,
({\bf b}-{\bf d}) DFT+DMFT features: ({\bf b}) spectral function $A(\bk,\omega)$, 
({\bf c}) Fermi surface and ({\bf d}) {\bf k}-integrated Bloch contribution $A_{\nu}(\omega)$ 
with characterization of dominance.
\label{fig:rhcr}}
\end{figure*}

Figure~\ref{fig:hdmft} displays the summarized DFT+DMFT spectral properties at room
temperature for the three heterostructures. A detailed discussion, also concenrning
stability issues and with extension to lower-temperature properties, can be found in 
Ref.~\cite{lec20}. The main result, common to all structural cases, may be extracted 
from the low-energy comparison of the $\bk$-resolved and the $\bk$-integrated data: while 
there are still QP-like dispersions visible in $A(\bk,\omega)$ at the Fermi level, the 
integrated spectra shows vanishing spectral weight at $\varepsilon_{\rm F}^{\hfill}$.
We coin this puzzling electronic state as correlation-induced semimetal (CIS), which is
obviously a result of the intriguing correlation scenario described above. Upon rising
obstruction of transport, the key Pd$(4d)$ electron can neither localize in real space 
(as in a Mott insulator) nor rest in a filled band (as in a band insulator). Hence it 
reduces the low-energy spectral weight as much as possible for an intact half-filled
band, resulting in the CIS state. Note that this finding is not an artifact of the
analytical continuation from Matsubara space to real frequencies, as the result is
comfirmed from both, maximum-entropy as well as Pad{\'e} methods~\cite{lec20}.
In some sense it amounts to a very strong reduction of the usual QP coherence scale of 
strongly correlated electrons, yet by still keeping the 'coherence' of the original dispersion. 
To our knowledge, such a rather exotic electronic state has not yet been reported in 
correlated matter and it awaits experimental verification.

Since the effective ${\cal A}$-site occupation is modified from the alternate-stacking 
architecture, the 
problem can also be pictured from an alternative viewpoint, namely from a formal change of 
the DFT filling factor $\alpha$ of Cr-$t_{2g}$ (cf. Fig.~\ref{fig:delcorr}b). The $\alpha$
values for the given heterostructures may formally be assigned by linear interpolation
of the values for PdCrO$_2$ ($\alpha=1/3$) and AgCrO$_2$ ($\alpha=1/2$). Accordingly,
the metallicity with respect to $\alpha$ is depicted in Fig.~\ref{fig:hdmft}j. The
transformation of the CIS states to the hidden-Mott state or the band-Mott insulator of 
the respective bulk compounds may be studied with different heterostructure layerings.
It is very likely from Fig.~\ref{fig:bmft}i and Fig.~\ref{fig:hdmft}i, that inbetween
the CIS and the band-Mott insulator, there is another metallic regime. Notably, hole
doping of the band-Mott-insulating state implements charge carrier into the more dominant
band-insulating part of AgCrO$_2$.
A doping $\alpha>1/2$ cannot be facilitated anymore by changing the ${\cal A}$-site 
TM$(4d)$ ion, but modifying the ${\cal B}$-site TM$(3d)$ might work. The regime $\alpha<1/3$
could in principle be reached by replacing Pd with Rh. But the Rh$^+$ ion with $4d^8$
configuration appears non-existing in known solid-state materials, and henceforth the 
stability of RhCrO$_2$ delafossite is very unlikely.
Still in a Gedankenexperiment, since also instructive for the understanding of the 
hidden-Mott state, we performed 
calculations for hypothetical RhCrO$_2$, using the lattice parameters and Hubbard $U$ of 
PdCrO$_2$. The results are shown in Fig.~\ref{fig:rhcr}. The DFT band structure at
lower energy exhibits the expected upward shiftings of the dispersions compared to the Pd 
compound. The filling of the Cr-$t_{2g}$ dominated bands is hence smaller, close to 
unity (i.e. $\alpha\sim 1/6$). 
Surprisingly, interactions still establish a near hidden-Mott state with a half-filled
Cr-$t_{2g}$ subshell. This underlines the significant coupling between the ${\cal A}$
layer and the CrO$_2$ planes. The fermiology becomes twofold in the Rh compound, associated
with a shrinking of the original warped hexagonal sheet and the appearance of novel
hole sheets around M (periodically arranged in a sixfold way). Note that $\Gamma$- and
M-sheet nearly touch in a Dirac-like crossing along $\Gamma$-M. The appearance of the 
M-sheet is not that surprising, since the corresponding dispersion with the maximum at M 
is observed just below the Fermi level in PdCrO$_2$ (cf. Fig.~\ref{fig:bmft}b). Finally,
the two-band picture is completed by observing two dominant LE-Rh contributions 
$A_{\nu}(\omega)$ close to $\varepsilon_{\rm F}^{\hfill}$. Furthermore, 
Fig.~\ref{fig:rhcr}d displays that the Cr-$t_{2g}$ dominated dispersions are not completely
transformed into Hubbard bands by interactions, but show reduced weight at the Fermi level.
In conclusion, if existing, RhCrO$_2$ would be a two-band metal with even stronger
entangling between Cr$(3d)$ and TM$(4d)$. Reaching such a two-band regime in metallic
delafossite, possibly by different doping/engineering routes of PdCrO$_2$, would be 
highly interesting.

\subsection*{In-plane alternations of PdCrO$_2$-AgCrO$_2$ type: 
Dirac(-like) states and emergent flat-band physics}
\begin{figure*}[t]
\begin{center}
\includegraphics*[width=16cm]{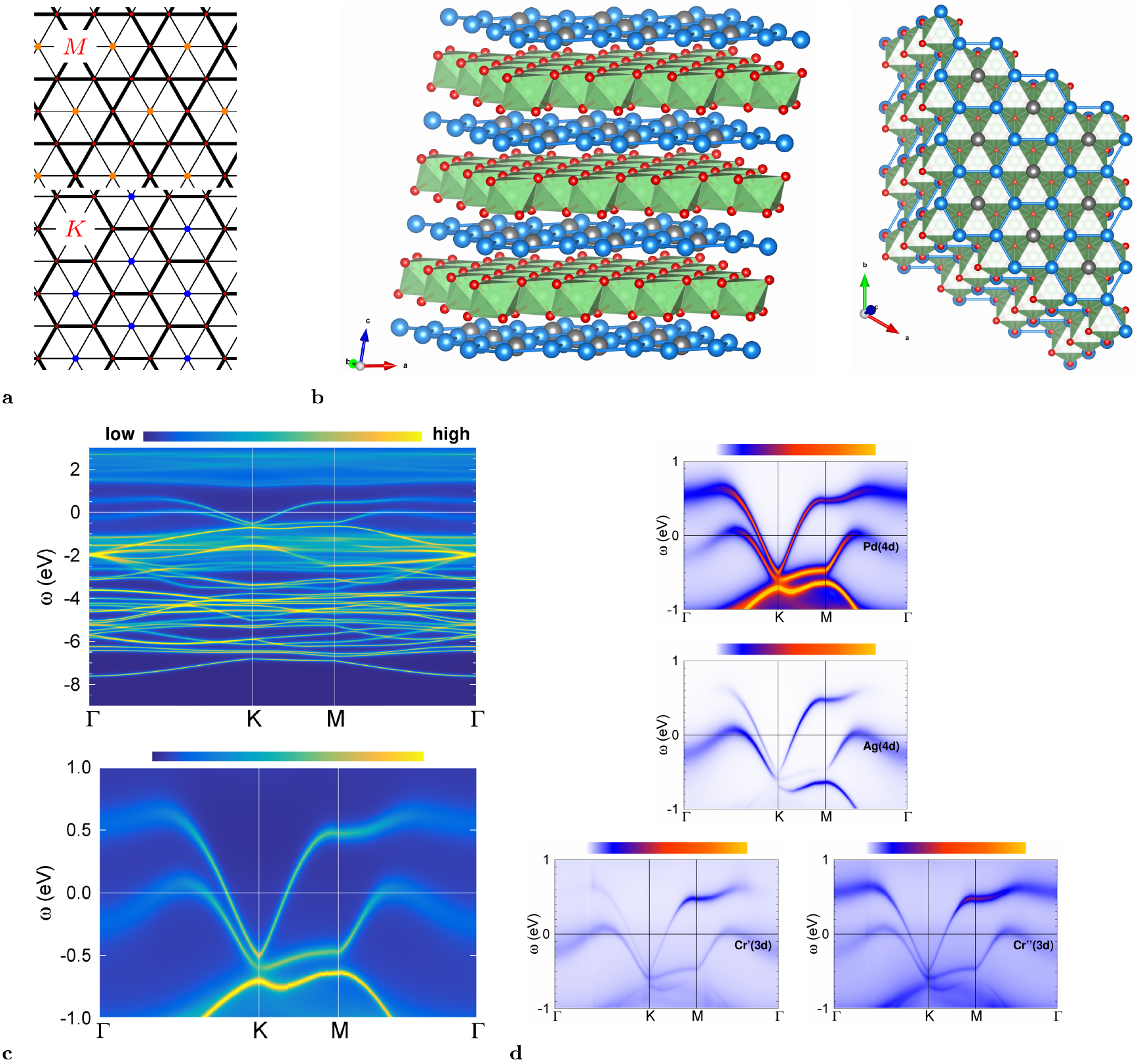}
\end{center}
\caption{{\bf Effective-honeycomb (Pd,Ag)CrO$_2$ delafossite}.
({\bf a}) In-plane design of the ${\cal A}$ layer via the natural triangular-lattice 
ordering instabilities of K-point (honeycomb) and M-point (kagom{\'e}) kind 
(from~\cite{boe12}). 
({\bf b}) Designed crystal structure from
understanding Pd(Ag) positions as active(blocking) sites; Pd: blue, Ag: grey,
O: red; Cr: green. Left: 3D view and right: view onto the effective honeycomb lattice
of Pd sites.
({\bf c},{\bf d}) Interacting spectral information from DFT+DMFT at
$T=193$\,K. ({\bf c}) Spectral function in (top) larger and (bottom) smaller energy window.
({\bf d}) Orbital-site content ('fatbands'), from top to bottom: Pd$(4d)$, Ag$(4d)$ as
well as both symmetry-inequivalent Cr$(3d)$, i.e. Cr'$(3d)$ and Cr''$(3d)$ (see text).
\label{fig:honey}}
\end{figure*}
Finally, let us push the limits of conceivable delafossite engineering even somewhat further,
by interpreting the metallic implication of Pd in PdCrO$_2$ and the band-insulating
implication of Ag in AgCrO$_2$ theoretically footlose. Instead of engineering PdCrO$_2$
'out of plane' from replacing Pd layers by Ag layers, one may imagine an 'in-plane'
alternation from replacing Pd sites by Ag sites in the periodically-repeated ${\cal A}$
layer. As a result, a novel natural-heterostructure delafossite emerges, but now with
a decorated ${\cal A}$ layer.
The viewpoint behind arises from picturing the ${\cal A}$ layer in hidden-Mott delafossite
as a canonical single-band triangular lattice at half filling, embedded in a Mott-insulating
background. By manipulating the features of this triangular lattice, a platform for 
studying correlation effects in such a Mott background may be generated. 
The simplest manipulations in this regard are given by the straightforward transformations 
of the original triangular lattice via the K- and M-point ordering instabilities, 
associated with the honeycomb (K) and the kagom{\'e} (M) lattice (see Fig.~\ref{fig:honey}a).
Realizing those lattices within a Mott background is exciting because they host 
Dirac-semimetallic and in the case of the kagom{\'e} lattice additionally flat-band
dispersions. The study of these dispersion features under the possible influence 
of strong correlations is a recent emerging research field in condensed 
matter, see e.g. Refs.~\cite{maz14,ye18,ghi20}.
\begin{figure*}[t]
\begin{center}
\includegraphics*[width=15.5cm]{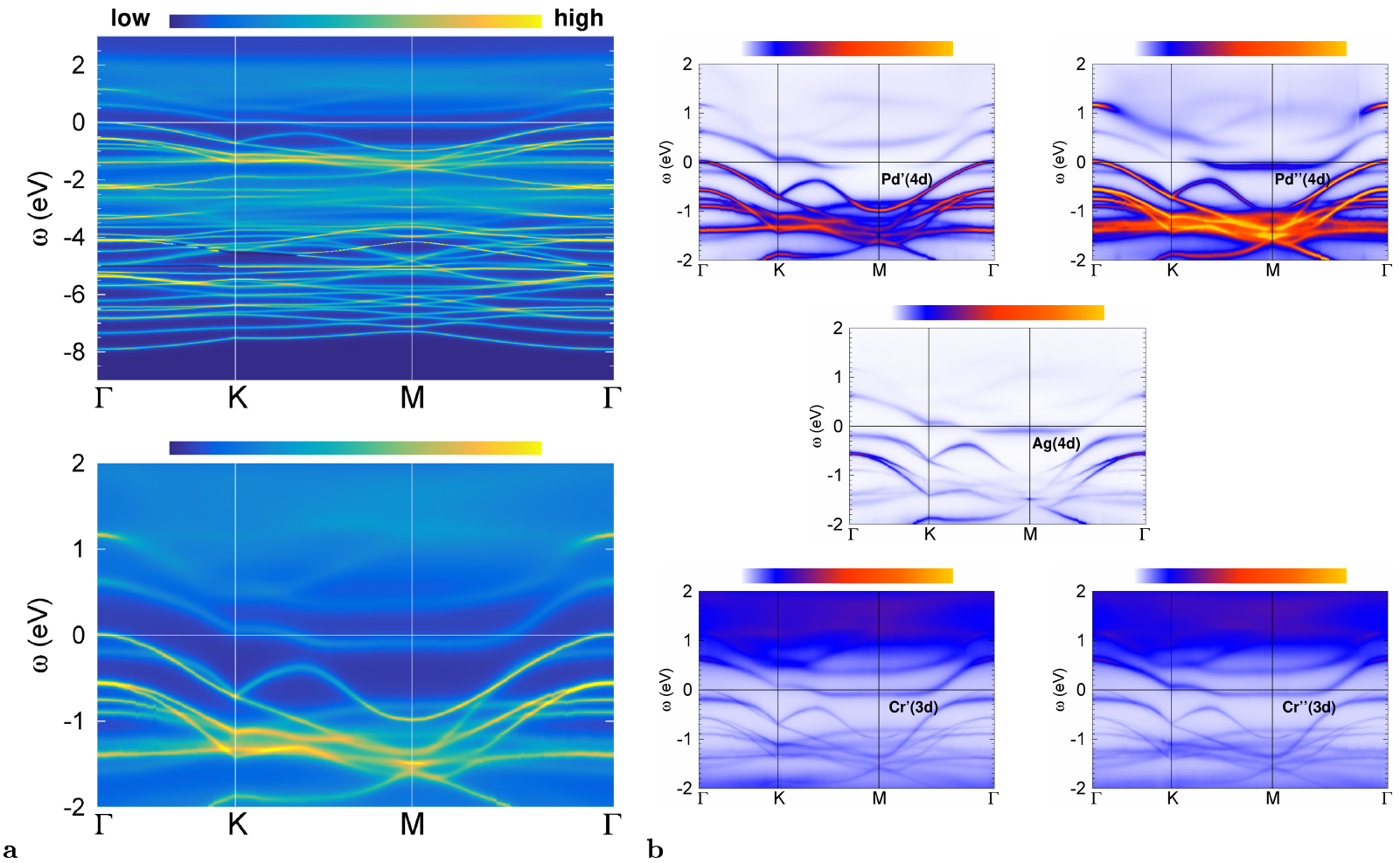}
\end{center}
\caption{{\bf Interacting spectral information for effective-kagom{\'e} (Pd,Ag)CrO$_2$ 
from DFT+DMFT}.
({\bf a}) Spectral function in (top) larger and (bottom) smaller energy window.
({\bf b}) Orbital-site content ('fatbands'), from top to bottom: 
both symmetry-inequivalent Pd$(4d)$, i.e. Pd'$(4d)$ and Pd''$(4d)$,
Ag$(4d)$ as well as both symmetry-inequivalent Cr$(3d)$, i.e. Cr'$(3d)$ and Cr''$(3d)$
(see text). All data for $T=193$\,K.
\label{fig:kago}}
\end{figure*}

To facilitate this reductions of the triangular lattice, the introduction of periodic 
blocking sites which ideally disconnect hopping processes, can be a promising route. 
This concept has been e.g. used to account for the effect of charge ordering onto transport 
properties in Na$_x$CoO$_2$~\cite{pie10,pei11}. From the nominal Ag$(4d)$ filled-shell in 
the CrO$_2$-based delafossites, we assume in the present context that Ag sites within 
the Pd layer may serve as such blocking sites. For sure, finite covalency will only realize 
a partial blocking, however, this may still be sufficient to 
mimic basic honeycomb- or kagom{\'e}-lattice features.
To realize a honeycomb(kagom{\'e}) lattice in that spirit within the ${\cal A}$ layer, 
one out of three(four) in-plane Pd sites has to be replaced by Ag. Note that though 
preparing such orderings in the lab will be surely demanding, a layer-by-layer growth might 
still be feasible from tailoring the layer stoichiometry. If the given Pd-Ag in-plane
orderings are thermodynamically stable (for given temperature, pressure, strain, etc.),
nature will take its course in realizing the periodic effective honeycomb/kagom{\'e}
pattern. For the honeycomb case, Fig.~\ref{fig:honey}b depicts the designed delafossite
structure. The DFT+DMFT calculations for both effective-lattice systems are again performed 
with corresponding linear interpolation of the known lattice parameters and the chosen 
Hubbard $U$ values for the bulk compounds. The honeycomb(kagom{\'e}) structure asks for 
a supercell of three(four) original formula units. Note that we utilitzed a somewhat
lower system temperature of $T=193$\,K for both structural cases, since from general
inspection of the spectral properties, the coherence scale for stable quasiparticles 
appears smaller for the given in-plane alternations compared to the out-of-plane
heterostructures.

In Fig.~\ref{fig:honey} we first show the spectral function of the effective honeycomb
structure. From graphene studies it is well known that the nearest-neighbor (NN) 
tight-binding electronic structure of the half-filled honeycomb lattice is semimetallic, 
with prominent Dirac dispersions (i.e. massless Dirac fermions) at the K point in reciprocal 
space~\cite{wal47}. The low-energy spectrum of the present effective lattice in the 
delafossite setting shows indeed some resemblance of this feature. First, the CrO$_2$
planes remain Mott-insulating upon the in-plane (Pd,Ag) structuring and there are two
Pd-dominated dispersion close to $\varepsilon_{\rm F}^{\hfill}$ (see Fig.~\ref{fig:honey}a).
A Dirac-like dispersion around K is indicated, yet shifted, with different filling and
different overall dispersion compared to graphene. Still, some blocking behavior
of Ag is realized, transforming the original PdCrO$_2$ low-energy dispersion in direction
towards the canonical honeycomb dispersion. The
site-orbital content in Fig.~\ref{fig:honey}b renders obvious that Ag$(4d)$ and Cr$(3d)$
have quite a weight on the twofold dispersion around the Fermi level, highlighting the
entangled nature. Note that there are two symmetry-inequivalent Cr positions, with 
Cr' mirroring the position of the blocking site within the Cr sublattice.

The NN tight-binding electronic structure of the kagom{\'e} lattice is known for its
flat-band feature at one side of the band edge, as well as for the Dirac dispersion
at 4/3(2/3) filling depending of the sign of the NN hopping (e.g.~\cite{maz14}). 
In the present case, the flat-band feature should appear at the upper band edge and 
thus the Dirac point at 2/3 filling. Figure~\ref{fig:kago} depicts the resulting
spectral function of effective-kagom{\'e} (Pd,Ag)CrO$_2$, and from a brief look the
canonical kagom{\'e} features are hard to decypher. The intriguing effect of 
correlations and only-partial Ag blocking render things hard to read. Yet after 
a closer look, and after also comparing with the non-interacting DFT states, 
the remains of the flat-band feature can be located around 1\,eV above the Fermi level.
Interestingly, the interactions in the delafossite structure seemingly transfer spectral 
weight from there towards $\varepsilon_{\rm F}^{\hfill}$. Close to $\Gamma$, a 
waterfall-like spectral signature may be observed. Hence a flat, but at $T=193$\,K rather
incoherent, low-energy feature ranging from K to M and halfways to $\Gamma$ emerges
(see Fig.~\ref{fig:kago}a). Its spectral content is dominated by the symmetry-equivalent
subclass Pd''$(4d)$, which collects two out of the three in-plane Pd sites of the 
supercell. But again, contribution from once more formally Mott-insulating Cr$(3d)$
is not negligible. The Dirac-dispersion feature at K from the canonical kagom{\'e} lattice
can also be identified at about -0.6\,eV. 
Thus albeit again the finite covalency of in-plane Ag does not allow for complete blockings,
and also the original half-filled nature of the Pd layer in PdCrO$_2$ is disturbed by
introducing Ag, the interplay of the hidden-Mott physics with model-kagom{\'e} features 
gives rise to interesting low-energy behavior. 

In conclusion, expectedly neither the effective-honeycomb nor the effective-kagom{\'e}
lattice realization in the modifed PdCrO$_2$ structure enables canonical
textbook dispersions at the Fermi level. But as a proof of principles, the in-plane 
engineering of delafossites may be a route to create nontrivial low-energy dispersions
which are subject to the puzzling layer-entangled correlation delafossite physics.

\subsection*{Experimental work on engineering correlated delafossites}
After discussing some theoretical designing ideas, we eventually take a quick look on 
the current status of designing and manipulating correlated delafossites in experiment.

The preparation of bulk single crystals has been demanding for quite some years, but
nowadays does not really pose a tough problem anymore for the known delafossites. They
are nowadays even used as electrocatalyst for hydrogen evolution reactions~\cite{gli19}.
However, layer-by-layer growth from e.g. pulsed-laser deposition (PLD) or molecular-beam 
epitaxy (MBE),
especially for the metallic systems, remained challenging until recently. But starting 
from about two years ago, the number of successful reports on grown metallic delafossites 
is increasing. For instance, thin-film preparation of 
PdCoO$_2$~\cite{har18,bra19,yor19,sun19} and PdCrO$_2$~\cite{ok20,wei20} has been 
reported by several groups. The transport properties of such films share in most cases
the exceptional high conductivity known from the bulk compounds, and device-oriented 
ideas have been proposed~\cite{har20}. Notably, in the ultra-thin limit of PdCrO$_2$ films 
grown on a single-layer of CuCrO$_2$, a significant increase of resistivity is 
reported~\cite{ok20}. 

Recent surface-sensitive studies on metallic delafossites are furthermore promising in
revealing details on the impact of correlations. Strong Rashba-like spin splitting in
CoO$_2$ and RhO$_2$ related delafossite surface states~\cite{sun17} and
itinerant ferromagnetism on the Pd-terminated (polar) surface of 
PdCoO$_2$~\cite{maz18,har20_2} have been observed.

Chemical doping and/or introduction of reasonable amounts of point defects through 
irradiation appears quite difficult in metallic delafossites~\cite{sun20-2}. Systematic
studies on substitutional doping are very rare~\cite{tan99}. Apparently, the high-purity
crystal state of these materials~\cite{mac17} renders it on the other hand hard to
implant a sizable amounts of point defects.

Nonetheless, the research activity on (metallic/correlated) delafossites is high and 
expected to further increase. Experimental investigations on specifically designed 
delafossites, e.g. along the lines of the theoretical propositions discussed in 
the previous section, are believed to become available soon.

\section*{Conclusions and outlook}
Delafossites are not new materials in condensed matter physics. They have been around 
for quite some time and also strong correlation phenomena have been discussed e.g. in the
context of insulating Cu-based compounds within the early days of high-$T_{\rm c}$ 
cuprate research. Yet the concise studies of high-purity metallic delafossites that
started about ten years ago brought them (back) into the spotlight of condensed matter 
research. Though known as the best-conducting oxides, they still share the
inherent property of hosting strongly correlated electrons common to most 
transition-metal oxides. But whereas in standard Mott materials such as e.g. V$_2$O$_3$,
YTiO$_3$ or La$_2$CuO$_4$ the competition between Coulomb repulsion and itinerancy
establishes an overall Mott-insulating state (or a metal-insulator transition with 
temperature), electron correlations in delafossites act much more subtle and their
characteristics can be quite elusive. In the extreme limit of CrO$_2$-based 
delafossites, they can give rise to layer-selective Mottness which is then coupled to 
the remaining metallic and/or band-insulating layers in an unusual way.

What makes delafossites so different from other oxides? The dumbbell-bond separation
between the ${\cal A}$ and ${\cal B}$O$_2$ layers leaves them rather
'freestanding', which enables a seemingly independent electronic behavior. As a result,
the scattering within the ${\cal A}$ layer and from there with the ${\cal B}$O$_2$ layers
is surprisingly small, for reasons which very details still need further exploration.
Notwithstanding, a delicate coupling between both layer types prevails, as e.g. 
displayed by the here presented results of the correlation-induced increase of Fermi
velocity in PdCoO$_2$ and the metal-to-metal transition with increasing $U$ in 
hidden-Mott PdCrO$_2$. In the general correlation context, the latter compound stands 
out and is undoubtly the most 'enigmatic' delafossite. But in standard measurements of
transport and spectroscopic behavior, there are (yet) no dramatic signs of its unusual
correlated nature. The recent photoemission study of Sunko {\sl et al.}~\cite{sun20},
which speculates about spin-charge separation in PdCrO$_2$, appears as an important
experimental step to face its correlated phenomenology. Albeit driving the system out of 
its 'comfort zone' seems necessary to unravel more facets of correlated electrons in
PdCrO$_2$.

Let us emphasize once more that the present review deals with a selective materials view 
on correlation effects in delafossites. There are various further members of this oxide 
family that deserve attention in this respect. For instance, the AgNiO$_2$ compound is
known to show challenging charge and spin ordering~\cite{waw07}, with possiblly increased
relevance of Hund's coupling physics~\cite{ral15}.
In view of magnetism, there is a broad literature on PdCrO$_2$ and related 
spin-active delafossites, e.g. 
Refs.~\cite{kan09,ars16,gha17,le18,sundan19,par20,kud20,kom20,sun20}, but in this
short review we focussed on the paramagnetic correlation aspects. The magnetic properties
are surely a relevant feature and based on correlations. However as often the case in
TM oxides, those properties seem secondary once strong correlations have been 
established. For instance, the 120$^\circ$ ordering at lower temperature in PdCrO$_2$
and AgCrO$_2$ may only set in after the Mott-critical behavior with strong local 
$S=3/2$ spin formation in the CrO$_2$ layers is realized. A deeper understanding of the 
underlying exchange mechanisms should therefore ask for a thorough account of the
correlated-electron states. It might also be conceivable that explicit local Coulomb 
interactions on the ${\cal A}$ site prove necessary for describing the fine details of 
delafossite magnetism. It this regard, it is understood that the spin degree of freedom 
may be crucial to access the delafossite many-body physics by experimental means.

As we stressed several times, the design aspect comes in naturally when engaging oneself
with delafossites. The bulk compounds are comparably simple in structure and there are
various different delafossites ranging from metallic to insulating. Plus, they may host
such intriguing phases as the hidden-Mott phase. Hence taking advantage of this 
potentially plethora of design possibilities is tempting and we tried to discuss some 
ideas in that direction in the present text. Besides magnetism, we also left out the
aspect of topology. The latter is very prominent in modern condensed matter research,
and could also play a vital role in engineered delafossites. One may surely imagine
architectures hosting nontrivial topological fermions within the layer-selective
Mott background of strongly correlated delafossites. A first minimal step towards such
scenarios has here been indicated by unveiling Dirac(-like) dispersions by design. 

To conclude, albeit delafossites have not yet entered the biggest stage of condensed matter
physics, they surely are a 'colorful' addition to the realm of correlated materials. It
is hoped that this brief review stimulates further theoretical and experimental research in
finding ways to unleash the so far mostly hidden nature of strong electronic correlations
in delafossites.

\section*{Acknowledgments}
The author is indebted to H. O. Jeschke, P. D. C. King, I. Krivenko, A. P. Mackenzie, 
L. Pourovskii, R. Richter, A. W. Rost, V. Sunko and P. Wahl for helpful discussions 
on various facets of delafossites as well as on computational aspects.
Financial support from the DFG LE-2446/4-1 project ``Design of strongly correlated
materials'' and the Psi-k network is acknowledged. Computations were performed at the 
JUWELS Cluster of the J\"{u}lich Supercomputing Centre (JSC) under project 
number hhh08. 

\section*{Data availability} 
All data are available from the corresponding author upon request.

\section*{Code availablity}
The code used to perform the DFT+DMFT calculations is available upon reasonable
request to the author.

\section*{Author contributions statement}
F.L. set up the problem, conducted the calculations, collected the data, 
analyzed the results and wrote the manuscript.

\section*{Additional information}
The author declares no competing financial interests.

\bibliographystyle{new}
\bibliography{bibdela}

\end{document}